\documentclass[aps,twocolumn,preprintnumbers,prd,superscriptaddress,nofootinbib,10pt,floatfix]{revtex4-2}

\usepackage{aas_macros}

\usepackage{graphicx,epsfig}
\usepackage{amsmath,amssymb}
\usepackage{lineno}
\usepackage{float}
\usepackage{fancyhdr}
\usepackage{xcolor}
\usepackage{subfigure}
\usepackage[english]{babel}
\usepackage{multirow}

\usepackage{hyperref}

\hypersetup{
 colorlinks=true,
 linktoc=all,
 linkcolor=red,
 citecolor=blue,
 urlcolor = blue}

\modulolinenumbers[5]
\newcommand{\dd}{\mathrm{d}} %diferencial
\newcommand{\ii}{\mathrm{i}} %unitat imaginaria
\newcommand{\eee}[1]{\mathrm{e}^{#1}} %exponencials
\renewcommand{\phi}{\varphi}

\newcommand{\sgn}{\mathrm{sgn}}

\newcommand{\bfeta}{\mbox{\boldmath{$\eta$}}}

\newcommand{\beq}{\begin{equation}}
\newcommand{\beqa}{\begin{eqnarray}}
\newcommand{\eeq}{\end{equation}}
\newcommand{\eeqa}{\end{eqnarray}}

\DeclareMathOperator*{\argmax}{arg\,max}

\hypersetup{
 colorlinks=true,
 linktoc=all,
 linkcolor=red,
 citecolor=blue,
 urlcolor = blue}

\begin{document}

\title{Probing cosmic strings via gravitational-wave lensing}

\author{Oleg Bulashenko}
\email{oleg@fqa.ub.edu}
\affiliation{Institut de Ci\`encies del Cosmos (ICCUB), Facultat de F\'{i}sica, Universitat de Barcelona, Mart\'i i Franqu\`es 1, E-08028 Barcelona, Spain}

\author{Nino Villanueva}
%\email{nino.villanueva@uv.es}
\affiliation{Departamento de Astronom\'{\i}a y Astrof\'{\i}sica, Universitat de Val\`encia, Dr. Moliner 50, 46100, Burjassot (Val\`encia), Spain}
\affiliation{IDAL, Electronic Engineering Department, ETSE-UV, University of Valencia, Avgda. Universitat s/n, 46100 Burjassot, Valencia, Spain}

\author{Roberto Bada Nerin}
%\email{roberto.bada-nerin@ligo.org}
\affiliation{Institut de Ci\`encies del Cosmos (ICCUB), Facultat de F\'{i}sica, Universitat de Barcelona, Mart\'i i Franqu\`es 1, E-08028 Barcelona, Spain}

\author{Jos\'e~A.~Font}
\affiliation{Departamento de Astronom\'{\i}a y Astrof\'{\i}sica, Universitat de Val\`encia, Dr. Moliner 50, 46100, Burjassot (Val\`encia), Spain}
\affiliation{Observatori Astron\`omic, Universitat de Val\`encia,  Catedr\'atico Jos\'e Beltr\'an 2, 46980, Paterna (Val\`encia), Spain}
%\email{j.antonio.font@uv.es}

%\date{\today, original date: June 2019}
\date{\today}

\begin{abstract}
We present a framework for detecting gravitational-wave signals lensed by cosmic strings (CSs), addressing a key gap in current searches. CSs, whose detection would provide a unique probe of high-energy physics and the early Universe, possess distinct topological and geometric features that require a dedicated search strategy. Our approach employs a full-wave transmission factor, expressed analytically via Fresnel integrals, which captures the characteristic diffraction and interference effects of the conical spacetime around a straight CS.
We contrast CS lensing with the well-studied point mass lens (PML) model, highlighting their fundamental differences: CS lensing depends on cosmological distances, string tension $\Delta$, and wavelength $\lambda$, and produces two non-amplified images set by the global conical geometry.
In contrast, PML lensing is governed by the distance-independent ratio $\sim M_{Lz}/\lambda$, where $M_{Lz}$ represents the redshifted mass of the lens, with image properties derived from the lens equation.
For BBH mergers lensed by CSs, we show that the waveforms exhibit a characteristic beating pattern or time-separated exact replicas. 
We derive a detectability bound on the string tension and, using Bayesian model selection, demonstrate that CS lensing is distinguishable from both unlensed and PML-lensed signals across a wide region of parameter space.

\end{abstract}

%\pacs{98.62.Sb, 98.80.Cq, 42.25.Fx, 42.25.-p}


\maketitle

%{
%\hypersetup{linkcolor=blue}
%\tableofcontents
%}
%\tableofcontents

\section{Introduction}

Cosmic strings (CSs) are one-dimensional topological defects predicted to arise during spontaneous symmetry-breaking phase transitions in the early Universe, within a wide range of quantum field theories, including grand unified theories and string theory~\cite{kibble76, vilenkin-shellard94, hindmarsh95, jeannerot-2003, copeland-kibble-2010}. Their detection would provide a unique observational window into high-energy physics and the early Universe, potentially validating fundamental theoretical models.
Advanced LIGO~\cite{Aasi:2014jea}, Advanced Virgo~\cite{Acernese:2014hva}, and KAGRA~\cite{KAGRA:2020tym} observatories form an international network of ground-based gravitational-wave (GW) detectors. Together, they constitute the LIGO-Virgo-KAGRA (LVK) Collaboration, which has recently undertaken its fourth observing run (O4) \cite{O4a-Intro}. The continuous improvement of detector sensitivity and global coordination among these observatories significantly enhance the prospects of detecting a broad range of GW sources, including potential signals from CSs.
The LVK Collaboration has been actively searching for GW signals emitted by CS loops and kinks~\cite{LIGO-CS-2017, LIGO-CS-2021,Abac_2025_cosm-hep-background_O4a}. When a network of CSs is assumed, their collective emission may contribute to the stochastic GW background~\cite{Damour-Vilenkin-2000,Damour-Vilenkin-2001,Damour-Vilenkin-2004,Siemens2007,Caprini2018}. In addition, several studies have explored the possibility of identifying individual GW bursts from CSs~\cite{Aurrekoetxea2023,Cuceu_2025_GW231123}.

The prospects for detecting CSs are expected to improve significantly with the advent of next-generation GW observatories. The planned space-based interferometer LISA will be sensitive to the millihertz frequency band~\cite{auclair-23,LISA-2024}, making it particularly well-suited to probe the stochastic GW background generated by CS networks~\cite{auclair2019lisa,boileau2021lisa,blanco2024lisa,dimitriou2025lisa}. 
Pulsar Timing Arrays (PTAs)
have recently reported evidence for a stochastic GW background in the nanohertz regime~\cite{kitajima2023pta,tan2025pta}. CSs are among the leading candidates to explain this signal \cite{Ellis2021,Blasi2021NANOGrav}.
Upgraded terrestrial detectors---including future versions of LIGO, Virgo, and KAGRA~\cite{O4a-Intro}, as well as third-generation observatories such as the Einstein Telescope (ET) \cite{punturo2010-ET,branchesi2025-ET} and Cosmic Explorer (CE)~\cite{evans2021-CE}---will achieve improved sensitivity to lower strain amplitudes and extend their reach across a broader frequency range. This enhanced bandwidth opens up the possibility of detecting GW bursts from cusps or kinks on cosmic string loops \cite{meijer2024-ET}, as well as from collapsing cosmic string loops themselves \cite{Aurrekoetxea2023}, while also complementing the low-frequency observations of LISA and pulsar timing arrays (PTAs).

While most current efforts focus on detecting GWs emitted directly by CSs, an alternative and complementary strategy is to search for CSs through their gravitational lensing effects on GWs. In this scenario, a GW emitted by a compact binary source located behind a CS is modified as it propagates past the string due to the conical geometry of the surrounding spacetime. This lensing can produce distinctive wave-optics effects, such as diffraction and interference, which become prominent when the GW wavelength is comparable to the characteristic scale of the lensing geometry.
Importantly, binary black hole (BBH) mergers are significantly more intense GW emitters than CSs, making them ideal background sources for lensing studies. As a result, current interferometric detectors such as LVK possess sufficient sensitivity to search for lensing signatures induced by CSs. In contrast, the direct detection of GWs emitted by CSs themselves—such as bursts or stochastic backgrounds—typically requires the enhanced sensitivity of future-generation observatories.

Lensing by CSs has been studied in the context of electromagnetic (EM) waves, where it produces pairs of images with identical brightness, separated by an angle set by the string tension~\cite{Sazhin2021review}. Observational strategies cover a wide range of frequency bands, including searches for anisotropies in the cosmic microwave background~\cite{hergt2017-CMB}, double-image signatures in galaxy surveys~\cite{Christiansen2011COSMOS}, and repeated images in fast radio bursts~\cite{Xiao2022FRBStrings}. 
For GWs, however, the phenomenology is even richer due to wave-optics effects: when the time delay between the two paths around the string is comparable to the GW period, characteristic interference patterns emerge~\cite{yamamoto03,suyama06,yoo13,pla-string16,pla-fresnel17,jung18}. Remarkably, even when only a single image is detectable, diffraction signatures can still betray the presence of the string~\cite{pla-string16,pla-fresnel17,jung18}.

The LVK Collaboration has conducted multiple searches for GW lensing~\cite{Hannuksela:2019kle,lensingO3a,lensingO3b}, primarily targeting strong lensing and microlensing effects caused by compact objects such as black holes, stars, or galaxies. 
Despite these efforts, no conclusive evidence for GW lensing has been reported~\cite{lensingO3b,janquart-23}. 
There exist, however, proposals 
that associate the unusually high masses inferred in some binary
systems with potential gravitational lensing magnification~\cite{diego-21,bianconi-22,canevarolo-24} (see also related discussions in~\cite{lensingO3a,janquart-GW230529}). 
More recently, the event GW231123, reported in~\cite{GW231123-LVK}, 
is noted as having the strongest observed support for point-mass lensing.
These studies, however, have not yet considered CSs as potential lenses. One reason is that GW lensing searches typically rely on matched filtering with template banks constructed from known compact lens models \cite{wright2025lensingflow}. CSs, with their distinct topological and geometric features, have not been incorporated into such templates, potentially leaving their lensing signatures undetected.

To address this gap, we introduce a comprehensive framework tailored for detecting gravitational lensing signatures from CSs. Central to our approach is a full-wave transmission factor that incorporates interference and diffraction effects from the conical spacetime geometry of CSs, providing a compact analytical expression in terms of Fresnel integrals readily available in scientific libraries such as {\sc Python’s} {\sc scipy}. This formulation enables the efficient construction of matched-filter templates and their straightforward implementation in data-analysis pipelines, laying the groundwork for targeted searches of CS-induced lensing effects in GW data. Focusing on the case of BBH mergers lensed by straight CSs, we characterize the resulting observational signatures, assess their detectability with current and future GW observatories, and examine potential mismodeling biases, matched-filtering strategies, and consistency tests to provide a quantitative evaluation of their impact on lensing detectability. Furthermore, we study the characteristic imprint left by CS lensing on detected GW and its discernibility from other, more common GW sources by computing the odds ratio PML lensed waveforms or unlensed waveforms.

This paper is organized as follows. 
In Sec.~\ref{sec-wave-eff}, we review the main aspects of wave-optics effects in gravitational lensing, distinguishing between compact lenses and cosmic strings, and introduce the key parameters that characterize each regime.
In Sec.~\ref{sec-transmission-factor}, we begin with the conical spacetime geometry and present the full-wave solution for the transmission factor, parametrized by two dimensionless quantities: the scaled GW frequency—set by the string tension—and the source offset relative to the line of sight. 
Unlike compact lenses, the thin-lens approximation does not apply to cosmic strings; no lens equation is required, as image positions are dictated directly by the global conical geometry and the string tension. 
Finally, we compare cosmic string lensing with the point mass lens, emphasizing the fundamentally different physical mechanisms and observational signatures.
Section~\ref{sec-lensing-cases} discusses the observational imprints of cosmic string lensing and their detectability.
In Sec.~\ref{sec-selection-bias}, we analyze potential selection biases, including matched filtering strategies and consistency checks, and provide a quantitative assessment of their impact on lensing detectability.
Finally, Sec.~\ref{sec-concl} summarizes our conclusions. Additional technical details and estimate of the detection rate 
are provided in the appendices.

\section{Wave-Optics Effects in Gravitational Lensing}
\label{sec-wave-eff}

Gravitational lensing is a phenomenon where the path of a wave is perturbed by the gravitational field of an intervening mass distribution. This effect occurs when the line of sight from a wave source to an observer is altered by the spacetime curvature caused by a foreground object. The effects of lensing can be broadly categorized into two main regimes, geometrical optics (GO) and wave optics (WO), with the relevant regime being determined by the relationship between the wavelength $\lambda$ and the characteristic scale of the lens.

\subsection{Lensing by compact objects}

For compact lenses, such as black holes or galaxies, two characteristic length scales are particularly relevant. The first one is the Schwarzschild radius of the lens, $R_{\rm S}=2GM_{Lz}/c^2$, which sets the fundamental scale of the lensing potential. Here, $M_{Lz} = M_L (1+z_L)$ represents the redshifted mass of the lens, $G$ stands for Newton's constant, and $c$ denotes the speed of light. 
The second one is the Einstein radius, $R_{\rm E}$, which characterizes the extent of the lensing region on the lens plane and is given by $R_{\rm E} = \sqrt{2R_{\rm S} \mathcal{D}}$.
Here $\mathcal{D}\equiv d_L d_{LS}/d_S$ represents the reduced distance, where $d_{LS}$ is the angular diameter distance between the source and the lens, and $d_L$ and $d_S$ are the angular diameter distances to the lens and source at redshifts $z_L$ and $z_S$, respectively \cite{schneider92}.
The Einstein radius scales as $R_{\rm E}\propto \sqrt{M_L}$ and is typically much smaller than the cosmological distances $d_{LS}$, $d_L$, and $d_S$. 
Lensing effects are strongest when the source, lens, and observer are nearly aligned within an angle of order $R_{\rm E}/d_L$, known as the Einstein angle.

Neglecting caustics, the GO approximation applies when $\lambda \ll R_{\rm S}$. In this regime, the observed signal is well described by the superposition of a small number of distinct rays (images), which can be analyzed using the stationary phase approximation. Conversely, when $\lambda \sim R_{\rm S}$, wave-optics effects such as diffraction and interference become important~\cite{deguchi86a,deguchi86b,nakamura98,nakamura99,takahashi03}. These effects are especially relevant for GWs, whose long wavelengths and high degree of spatial coherence over cosmological distances make them more sensitive to WO phenomena than EM waves.

A useful parameter to quantify wave effects is the Fresnel number, defined from the ratio of the Einstein radius to the Fresnel scale $R_{\rm F} = \sqrt{\lambda \mathcal{D}}$. It can be written as~\cite{deguchi86b,BU-JCAP-21}
\begin{equation}
N_{\rm F} = \frac{R_{\rm E}^2}{R_{\rm F}^2} =\frac{2R_{\rm S}}{\lambda}.
\label{eq:NF_M}
\end{equation}
$N_{\rm F}$ represents the number of Fresnel zones contained within the Einstein ring that contribute to the lensing. The GO approximation holds when $N_{\rm F}\gg 1$, while WO becomes relevant for $N_{\rm F}\sim 1$. Notably, for a compact-mass lens, Eq.~\eqref{eq:NF_M} shows that $N_{\rm F}$ is independent of the source and lens distances.

\subsection{Lensing by Cosmic Strings}

The WO regime is especially important when the lens is a linear topological defect, such as a CS. A key parameter of a CS is its dimensionless tension, $\mu_G\equiv G\mu/c^2$, where $\mu$ is the mass per unit length. 
This parameter is directly related to the effective energy scale of the underlying theory describing string formation. Because the curvature of a CS is confined to its core, the spacetime around a straight string is locally flat, producing no gravitational attraction in its vicinity. On larger scales, however, the global conical topology gives rise to lensing effects with distinctive observational signatures \cite{vilenkin-shellard94,hindmarsh95}. For EM radiation, a light ray passing near the string is deflected by a constant angle $\Delta = 4\pi \mu_G$, regardless of the impact parameter. This produces double images of a background source located behind the string, with an angular separation of approximately $2\Delta$~\cite{vilenkin81,vilenkin84,gott85}. Since a straight string has no focusing effect, the two images have equal brightness.

In the case of GWs, whose long wavelengths remain coherent over cosmological distances, wave effects play a central role. 
When the time delay between the two images is comparable to the wave period, the GW passing on opposite sides of the string interfere with itself, leaving a measurable imprint in the lensed waveform~\cite{yamamoto03}. Moreover, even in situations where only one image is visible---while the second is obscured—interference due to diffraction around the string can still occur~\cite{suyama06,yoo13,pla-string16,pla-fresnel17,jung18}. This phenomenon can be interpreted within the geometrical theory of diffraction as the interference between a direct GW path and a diffracted component emanating from the location of the string~\cite{pla-string16,pla-fresnel17}.

If the lens is a linear topological defect such as a CS, the Einstein radius is
$R_{\rm E} = \mathcal{D}\Delta$, and the corresponding Einstein angle is $R_{\rm E}/d_L = (d_{LS}/d_S)\Delta$, where $\Delta$ is determined by the string tension. 
In the relevant regime, $d_S \sim d_{LS} \gg d_L$, 
the Einstein angle reduces to a constant equal to $\Delta$. The Fresnel number for CS lensing is then

\begin{equation}
    N_{\rm F} = \frac{\mathcal{D}^2\Delta^2}{R_{\rm F}^2} = 
    \frac{\mathcal{D}}{\lambda}\, \Delta^2.
    \label{eq:NF_S}
\end{equation}

In contrast to compact lenses, the Fresnel number in the CS case depends explicitly on the distances between source, lens, and observer~\cite{pla-string16,pla-fresnel17}.

%%%%%%%%%%%%%%%%%%%%%%%%%%%%%%%%%%%%%%%%%%%

%------------------------------------------------------
\section{The transmission factor}
%------------------------------------------------------

\label{sec-transmission-factor}

\subsection{Conical spacetime}
Suppose that the string (referred to as the ``lens'') is positioned very 
close to the line of sight between the source of the GW and the detector. 
For simplicity, we assume that the distance $d_{LS}$ between the GW source and the string is much longer than $d_L$ -- the distance between the string and the detector. In this case the incident wave is just a plane wave and the reduced distance $\mathcal{D}\approx d_L$. 
Additionally, the string is considered to be orthogonal to the line of sight.
This assumption does not preclude the application of our analysis to the 
possibility of inclination.

\begin{figure}[t]
\includegraphics[width=0.90\columnwidth]{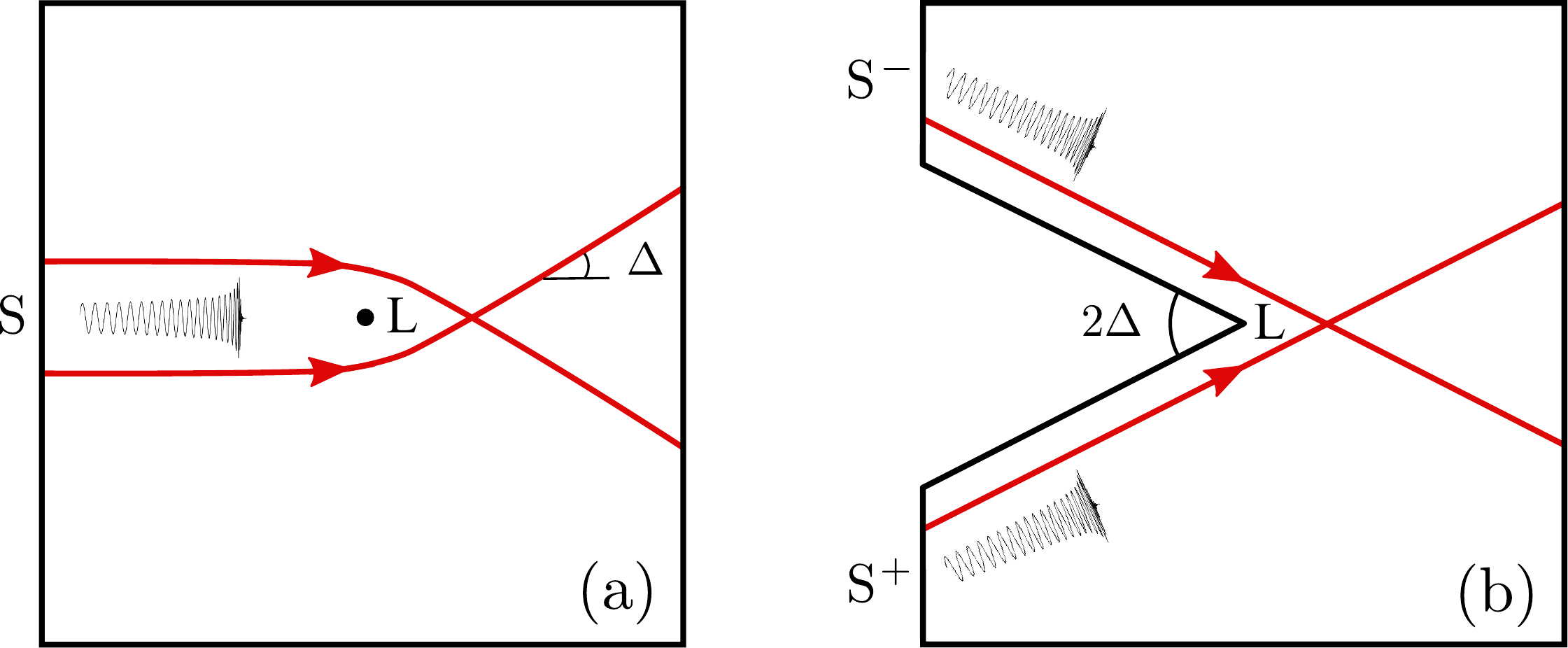}
\caption{
Schematic diagram illustrating wave propagation in a conical space at the $z=0$ plane: 
(a) The source $S$ emits a GW. Two representative geodesics (depicted in red) pass on opposite sides of the string $L$ and are deflected by the angle $\Delta$; 
(b) An equivalent space with a deficit angle of $2\Delta$ and two image sources, $S^-$ and $S^+$. The geodesics are straight lines in this configuration.}
\label{fig:geom}
\end{figure}

Consider the spacetime metric for a static cylindrically symmetric cosmic string lying along the $z$ axis \cite{vilenkin81,vilenkin84,gott85} 
\begin{equation}
\dd s^2= -c^2\dd t^2 + \dd r^2 + (1-4\mu_G)^2 r^2\dd \phi^2 + \dd z^2,
\label{eq:metric-phi}
\end{equation}
where $\mu_G$ is the string tension.
The topology of this space is nontrivial in the sense that it is locally flat everywhere except at the origin where the curvature has a singularity.
This singularity is precisely the reason why null geodesics are all
deflected by the same angle,
\begin{equation}
\Delta = 4\pi \mu_G\,,
\label{eq:delta}
\end{equation}
determined by the string tension \cite{vilenkin81,vilenkin84} 
[see Fig.~\ref{fig:geom}(a)].
Since geodesics passing on opposite sides of the string eventually cross, one
should expect interference and diffraction effects.
With a new angular coordinate $\theta=(1- 4\mu_G) \phi$,
the metric \eqref{eq:metric-phi} takes a Minkowskian form
\begin{equation}
\dd s^2= -c^2\dd t^2 + \dd r^2 + r^2\dd \theta^2 + \dd z^2,
\label{eq:metric-Mink}
\end{equation}
with a wedge of angular size $2\Delta$ removed and the two faces of the wedge  identified \cite{vilenkin81,vilenkin-shellard94}.
In this space, the angular coordinate $\theta$ spans the range $2\pi - 2\Delta$,
and the wave source $S$ is doubled into two images $S^-$, $S^+$ which are 
located on the faces of the wedge [see Fig.\ \ref{fig:geom}(b)]. Each image 
source emits an identical GW signal propagating along a straight-line geodesic 
(see \cite{pla-string16} for further details).

\subsection{Full-wave solution}
\label{subsec:full-wave}

An advantage of the geometry given in \eqref{eq:metric-Mink} is that, for each time-harmonic incident field, it allows the construction of a lensed solution 
from the canonical Sommerfeld's solution of wave diffraction on a half-plane screen. 
That solution is mathematically exact, in contrast to the Kirchhoff's formulation, which is only approximate \cite{sommerfeld}.
Following \cite{pla-string16}, the wave field at the observation point $(r,\theta)$ can be written as a superposition of two penumbra contributions
\begin{equation}
U =  \mathcal{F}(u^+) \, \eee{\ii kr\cos(\Delta+\theta)}
+ \mathcal{F}(u^-) \,\eee{\ii kr\cos(\Delta-\theta)},
\label{eq:u-diff}
\end{equation}
where $k=2\pi f/c$ is the wavenumber,
$r$ is the radial coordinate (equal to $d_L$ in our setup), $\theta=(1-\Delta/\pi)\,\phi$ is the angular coordinate, and 
$u^\pm \equiv\sqrt{2kr}\sin\,[(\Delta\pm\theta)/2]$.
The function
\begin{equation}
\mathcal{F}(u) = \displaystyle{ \frac{e^{-\ii\pi/4}}{\sqrt{\pi}} 
\int_{-\infty}^u \eee{\ii s^2}\dd s}
\label{eq:Fresnel-int}
\end{equation}
is the Fresnel integral, normalized such that $\mathcal{F}(\infty)=1$.~\footnote{The integral~\eqref{eq:Fresnel-int} can be evaluated numerically using the Fresnel sine and cosine functions, $S(x)$ and $C(x)$, available in {\tt scipy.special.fresnel}. In particular, one may use the relation $\mathcal{F}(u) = [(1+\ii)/2 + C(\sqrt{2/\pi}\,u) + \ii\, S(\sqrt{2/\pi}\,u) ]/\sqrt{2\ii}$.}
Each term in the solution \eqref{eq:u-diff} corresponds to one of the two images positioned at $\theta=\pm(\pi-\Delta)$. In this geometry, the line of sight corresponds to 
$\theta=0$. 
As demonstrated below, Eq.~\eqref{eq:u-diff} encompasses both geometric optics and wave effects (diffraction).
In the no-lens limit ($\Delta=0$, $\mathcal{F}(0)=1/2$), the wave field 
reduces to the unlensed plane wave $U_0=\eee{\ii kr \cos\theta}$.
The transmission factor\footnote{The transmission factor is also referred to as the amplification factor~\citep{takahashi03}.
We prefer the term ``transmission'', as the signal passing through a cosmic string is not truly amplified---instead, it splits into two images and is modified by diffraction effects.}
$F=U/U_0$ can now be determined 
from Eq.~\eqref{eq:u-diff} and expressed as a function of two variables, the 
GW frequency $f$ and the polar angle $\theta$. 
Considering that $\Delta\ll 1$ and introducing the normalized angle $y \equiv \theta/\Delta$, we finally obtain the full-wave transmission factor,
\begin{eqnarray}
F(f) &=& 
\mathcal{F}(\sigma^+ \sqrt{2\pi f t_d^+}) \,\, 
\eee{-2\pi \ii f t_d^+} 
\nonumber\\ 
&+&\mathcal{F}(\sigma^- \sqrt{2\pi f t_d^-}) \,\, \eee{-2\pi \ii f t_d^-},
\label{eq:F-f-full}
\end{eqnarray}
where $t_d^\pm$ are the time delays associated with each image, given by
\begin{equation}
t_d^\pm = t_{\Delta} \,(1\pm y)^2 /2,  \qquad 
t_{\Delta} = \frac{\chi_L}{c} \Delta^2,
\label{eq:t_d}
\end{equation}
and $\sigma^\pm \equiv \sgn(1 \pm y)$ are sign functions that take the value $+1$ in the illuminated region and $-1$ in the shadow.
Here $\chi_L=(1+z_L)\,d_L$ denotes the comoving distance to the lens, where $z_L$ represents the redshift~\cite{schneider2006book}.
In these notations the angles $-1<y<1 $ correspond to the double imaging range.
It is worth noting that the frequency scales inversely with $t_{\Delta}$ and the product $ft_{\Delta}$ is equal to $N_{\rm F}$ -- the number of Fresnel zones inside of the Einstein ring contributing to the lensing [compare with Eq.~\eqref{eq:NF_S}].

Although Eq.~\eqref{eq:F-f-full} is self-consistent and enables the computation of the transmission factor over the full range of parameters, it is helpful to examine certain limits---such as the GO and line-of-sight cases. These limits simplify the general expressions into more analytically tractable forms, making the physical interpretation more transparent. 

\subsection{High frequency limit}

At high frequencies, $ft_{\Delta}\gg 1$, the Fresnel integrals can be approximated using the asymptotic expansion
\cite{sommerfeld},
\begin{equation}
\mathcal{F}(u) = \mathcal{H}(u) -
\frac{\eee{\ii\pi/4}}{2\sqrt{\pi}}\, \frac{\eee{\ii u^2}}{u}\,
+ O(u^{-3}) \,,
\label{eq:expansion}
\end{equation}
with $\mathcal{H}(u)$ being the Heaviside step function. 
Substituting into Eq.~\eqref{eq:F-f-full}, we obtain a simplified expression for the transmission factor in the geometrical theory of diffraction (GTD) approximation~\cite{pla-string16}:
\begin{eqnarray}
F_{\rm GTD}(f) &\approx& 
h^+ \,\eee{-2\pi \ii f  t_d^+} + h^- \,\eee{-2\pi \ii f  t_d^-}
+ \frac{\tilde{A}}{\sqrt{ft_{\Delta}}} \nonumber \\ 
&\equiv& F_{\rm GO} + \frac{\tilde{A}}{\sqrt{ft_{\Delta}}}
\label{eq:F-GTD}
\end{eqnarray}
which contains explicit contributions from GO and from diffraction. 
The first two terms represent the lensed GO rays deflected on opposite sides of the string, which acquire a relative time delay of $t_d^+ - t_d^-$. 
The Heaviside functions, $h^\pm \equiv \mathcal{H}(1\pm y)$, ensure that each ray contributes only within its corresponding illuminated region.
Importantly, the amplitude of each GO ray is unity, implying that the rays are not amplified. Moreover, no Morse phase shift arises, since both paths correspond to minima of the travel time. This contrasts with lensing by compact objects, such as PML (see Sec.~\ref{compar-PML}), where the second image corresponds to a saddle point and thus acquires a Morse phase shift, while the amplitudes of the two images generally differ.
The last term in Eq.~(\ref{eq:F-GTD}) decaying with frequency $\sim O(f^{-1/2})$ represents the leading order contribution of the diffracted field.
It corresponds to a cylindrical wave emanating from the string, with a $y$-dependent diffraction coefficient that also determines its phase shift~\cite{pla-string16}
\begin{equation}
\tilde{A} = -  \frac{\eee{\ii \pi/4}}{\pi} \frac{1}{1-y^2}.
\label{eq:Diff}
\end{equation}

\begin{figure}[t]
\includegraphics[width=0.95\columnwidth]{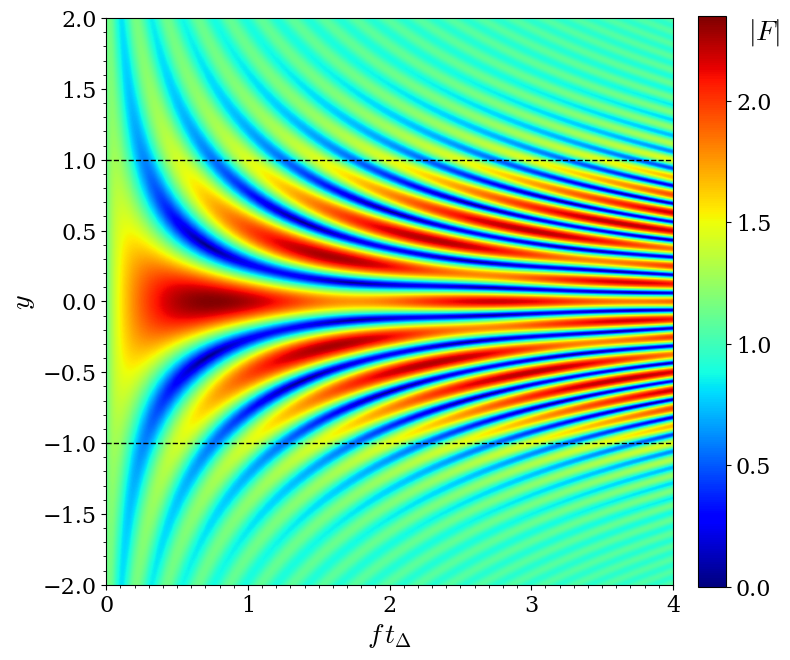}
\caption{Density plot of the transmission factor $|F|$ for a cosmic string lens as a function of the rescaled frequency $ft_{\Delta}$ and source position $y=\theta/\Delta$. The dashed lines mark the boundary of the double-imaging region ($|y|<1$).}
\label{fig:F-string-full}
\end{figure}

\begin{figure}[t!]
\includegraphics[width=0.95\columnwidth]{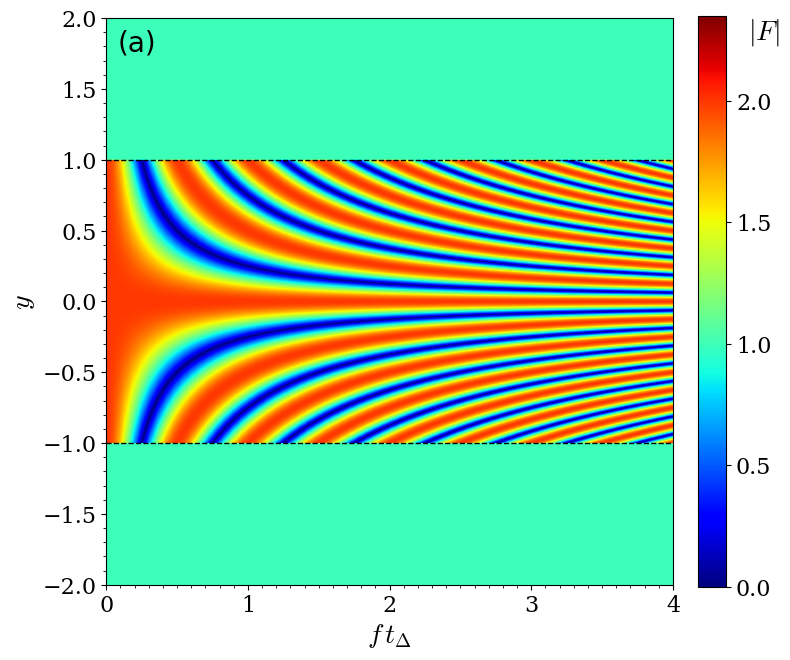}\\
\vspace*{2ex}
\includegraphics[width=0.95\columnwidth]{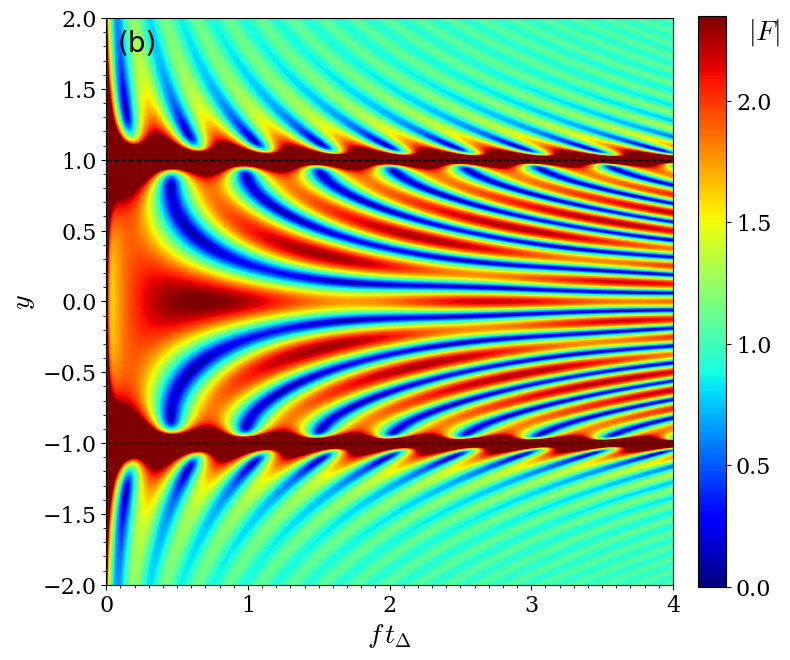}
\caption{Approximations to the transmission factor $|F|$ from Fig.~\ref{fig:F-string-full}: (a) geometrical optics (GO); (b) geometrical theory of diffraction (GTD), which extends GO by including an additional diffracted contribution [Eq.~\eqref{eq:F-GTD}].}
\label{fig:F-approx}
\end{figure}

Since the transmission factor $F$ depends on two variables---the frequency $f$ and the observation angle $y$---we illustrate the absolute value of $|F|$ in the two-dimensional parameter space $(f,y)$. The full-wave solution~\eqref{eq:F-f-full}, shown in Fig.~\ref{fig:F-string-full}, can be directly compared with the GO and GTD approximations in Fig.~\ref{fig:F-approx}. Wave effects manifest as interference fringes in the parameter space and, within the double-imaging region ($|y|<1$), are well captured by the interference of the two GO rays [Fig.~\ref{fig:F-approx}(a)].  
Outside this region ($|y|>1$), however, the GTD approximation—with its additional diffracted contribution—is required to reproduce the interference fringes [Fig.~\ref{fig:F-approx}(b)]. It is also needed to explain the local maxima along the antinodal lines [Fig.~\ref{fig:F-approx}(b); see also the discussion of Fig.~\ref{fig:noadal-lines} in Appendix~\ref{appendixA}].  
Overall, GTD offers an accurate description across the parameter space, except near the boundary $|y|\approx 1$, where the diffraction term diverges, and in the low-frequency limit.

\subsection{Interference fringe}

To assess the observational signatures of GW diffraction by cosmic strings, we focus on the double-imaging region, $|y|<1$, where interference effects are strongest and of greatest observational relevance. Our goal is to characterize the interference fringe pattern in the transmission factor, determine its frequency scale, and compare it with the typical scales of GW sources. 
In this region, Eq.~(\ref{eq:F-GTD}) takes the form  
\begin{equation}
F_{\rm GTD}(f) =2 \,\eee{-\ii \pi (1+ y^2) f\,t_{\Delta}} \cos(2\pi f\,t_{\Delta} \,y)
+ \frac{\tilde{A}}{\sqrt{ft_{\Delta}}}.
\label{eq:F-2i-reg}
\end{equation}
From this expression, it follows that, for any fixed $y\neq 0$, the  transmission factor exhibits regular oscillations with evenly spaced maxima and minima (see Fig.~\ref{fig:F-string-full}).
These oscillations originate from the crossing of nodal and antinodal lines---constant-phase contours defined by the interference of GO rays (for details see Fig.~\ref{fig:noadal-lines} in Appendix~\ref{appendixA}). These lines create the fringe pattern in the transmission factor and, consequently, in the GW lensed waveform \cite{jung18}.

We can determine the fringe spacing by analyzing the first GO term in Eq.~\eqref{eq:F-2i-reg}: 
\begin{equation}
|F_{\rm GO}(f)| \approx 2\, |\cos(2\pi ft_{\Delta} \,y)|.
\end{equation}
Thus, the fringe spacing is uniform in frequency, with a characteristic separation given by,  
\begin{equation}\label{eq:f_delta}
f_{\Delta} = \frac{1}{2 t_{\Delta} \,y} = \frac{c}{2\chi_L\Delta^2 \,y}\,.
\end{equation}
This result agrees with previous findings \cite{suyama06,pla-string16,jung18}.
Using Eq.~\eqref{eq:t_d}, we estimate the characteristic time delay: 
\beq
t_{\Delta} = 1.029\, s \left(\frac{\chi_L}{100\, {\rm Mpc}} \right) \left(\frac{\Delta}{10^{-8}} \right)^2,
\label{t_Delta}
\eeq    
which leads to the frequency scale of the GW fringes, 
\beq
f_{\Delta} \approx 48.6\, {\rm Hz} \left(\frac{1}{y} \right)
\left(\frac{100\, {\rm Mpc}}{\chi_L} \right) \left(\frac{10^{-9}}{\Delta} \right)^2.
\label{eq:fringes}
\eeq
To determine whether wave effects induced by string lensing are observable, these scales should be compared with the typical duration and frequency content of the GW source---in our case, an inspiraling binary.

\subsection{Comparison with Point Mass Lens}
\label{compar-PML}

Next, we consider an isolated  point mass as the lens for comparison. In gravitational lensing, the point-mass lens (PML) model is appropriate when the physical size of the lens is much smaller than the Einstein radius, as in the case of black holes, dense dark matter clumps, or similar compact objects. Due to its simplicity, the PML model has been widely used in the literature to interpret both EM \cite{deguchi86a,deguchi86b,schneider92} and GW lensing \cite{nakamura98,nakamura99,takahashi03,matsunaga06,christian18,diego19,liao19,cheung20,cremonese21a,seo21,wright21,BU-JCAP-21,caliscan22,ali22,tambalo22a,bondarescu22,LiuA-Wong23,savastano23,mishra23a,mishra23b,wright24,chen24,villarrubia24,chan-2024-Detectability_lensing_matched_filter,Roberto:2025}. 
It is currently employed in various analysis pipelines, such as {\sc Gravelamps} \cite{wright21}, {\sc Gwmat} \cite{mishra23b}, {\sc GLoW} \cite{villarrubia24}, and deep-learning autoencoders \cite{Roberto:2025}, among others, to search for microlensing signatures. In addition, model-independent techniques are also being developed for this purpose \cite{chakraborty2025-GLANCE}.

The Einstein radius $R_{\rm E}$ representing the characteristic length scale on the lens plane is typically much smaller than the cosmological distance $\mathcal{D}$. 
This enables the lens mass to be projected onto a lens plane. In the thin-lens approximation, GWs propagate freely outside the lens, interacting only with a two-dimensional gravitational potential at the lens plane, where the lensing effect is ultimately captured in the transmission factor $F$. The latter is determined by the Fresnel-Kirchhoff diffraction integral
\footnote{In our implementation, we take the complex conjugate---replacing $\ii$ with $-\ii$---to match the Fourier transform convention used in the Python libraries of the LVK Collaboration~\cite{lalsuite-fourier}.}
across the lens plane \cite{schneider92}
\beq
F(f,\mathbf{y}) = -\ii f\, t_M
\iint e^{\ii \,2\pi f\, t_d (\mathbf{x},\mathbf{y})} \, \dd^2\mathbf{x},
\label{F_w}
\eeq
where $\mathbf{x}$ denotes a two-dimensional coordinate vector in the lens plane, expressed in units of $R_{\rm E}$, while $\mathbf{y} = \bfeta \,d_L/(R_{\rm E} d_S)$ represents the normalized projection of the source position $\boldsymbol{\eta}$ (in physical units) from the source plane onto the lens plane~\cite{takahashi03}. 
The lensing time delay function is given by
\beq
t_d(\mathbf{x},\mathbf{y}) = t_M \, \left( \frac{1}{2}\, | \mathbf{x}-\mathbf{y} |^2
- \psi(\mathbf{x}) 
%+ \phi_m(\mathbf{y}) 
\right)\,,
\label{t-delay}
\eeq
where the characteristic lensing time scale $t_M$, expressed in physical units, is given by
\beq
t_M = (1+z_L) \,\frac{R_{\rm E}^2\,d_S}{c\,d_L d_{LS}} 
\approx \; 1.97 \times 10^{-5} \; \left(\frac{M_{Lz}}{M_\odot}\right) \,{\rm s}\,,
\label{eq:tM}
\eeq
and is proportional to the redshifted lens mass $M_{Lz} = M_L (1 + z_L)$.
The lensing properties are encoded in the lensing potential $\psi(\mathbf{x})$. 
For the PML, characterized by $\psi(\mathbf{x})=\ln{|\mathbf{x}|}$, the diffraction integral in Eq.~\eqref{F_w} admits an analytical solution \cite{schneider92,deguchi86a},
\beq
|F|=\displaystyle{
\left( \frac{2 \pi^2 \nu}{1-e^{-2 \pi^2 \nu}} \right)^{1/2}
 \left| _1F_1 ( \ii \pi \nu; \,1; \,\ii \pi \nu y^2 ) \right| },
\label{eq:abs_F}
\eeq
where $\nu \equiv f t_M$ is the dimensionless frequency, defined as the product of the GW frequency $f$ and the characteristic lensing time, $t_M$.
The function ${}_1F_1(a,b,z)$ denotes the confluent hypergeometric function.

At high frequencies, computing the transmission factor \eqref{eq:abs_F} becomes numerically expensive. In this GO limit the dominant contribution arises from two well-defined images of the source \cite{schneider92,nakamura99,takahashi03}, 
corresponding to two stationary points of the time delay function \eqref{t-delay}. 
The positions of the images $\mathbf{x}_j$ are determined by the lens equation, which for the PML model is given by $\mathbf{y}=\mathbf{x}-\mathbf{x}/|\mathbf{x}|^2$ \cite{schneider92}.
For $\mathbf{y}=(y,0)$, with $y>0$, the two images on the lens plane $\mathbf{x}_{1,2}=(x_{1,2},0)$ are determined by
$x_{1,2}=(y\pm \sqrt{y^2+4})/2$, which leads to the following formula for the transmission factor \cite{takahashi03}:
\beq
F_{\rm GO}= \left(\sqrt{\mu_+} + \sqrt{\mu_-} \,e^{2 \ii \alpha} \right) e^{\ii \phi_1}\,.
\label{F_GO_PML}
\eeq
Here, the magnification for each image, expressed as $\mu_{\pm}=(v+v^{-1}\pm 2)/4$ with $v \equiv y/\sqrt{y^2+4}$, is ultimately a function of $y$ alone, while the phase 
$\alpha = \pi f t_M \tau_{21} - \pi/4$, where 
$\tau_{21} = 2v/(1-v^2) + \ln [(1+v)/(1-v)]$, 
also depends on frequency and lens mass. 
The function $t_M \tau_{21}$ represents the time delay between the two images, with $\phi_1$ denoting the phase of the first image, taken as a reference.  
Under the ``close alignment'' condition ($y \lesssim 0.5$), this delay can be approximated as ~\cite{BU-JCAP-21}
\beq
\Delta t_{21} \approx 2 y\, t_M.
\label{eq:t21-PML}
\eeq

PMLs and CSs, while both capable of gravitational lensing, exhibit fundamental distinctions in their nature and lensing mechanisms. A PML is characterized by its compact mass localization in space, allowing its lensing effects to be effectively treated within the thin-lens approximation. Conversely, a CS represents an extended mass distribution in one dimension, thus being non-compact. Its lensing arises from the global conical structure of spacetime, rendering the strict thin-lens approximation inapplicable as the deflection of passing waves is a consequence of spacetime topology rather than a localized two-dimensional potential. The image formation process also diverges significantly. For a PML, the two lensed images are determined by a lens equation, with magnifications and phase shifts intricately linked to the lensing potential. In contrast, CSs do not adhere to a lens equation in the same manner. Under close alignment, they produce two images, exact copies of the original signal with equal ``magnification", characterized by a fixed deflection angle, $\Delta = 4\pi \mu_G$, with their positions dictated by the conical spacetime geometry.
Depending on the string tension, these two images may overlap and interfere, causing a beating pattern (weak lensing), or can be separated (strong lensing), as will be described in the next section. For larger angle offsets, only one image is observed, which is affected by diffraction. 

Consequently, PMLs and CSs affect passing GWs differently, leading to distinct manifestations of wave effects. For a PML, the key dimensionless quantity governing these effects is 
$2R_{\rm S}/\lambda$, notably independent of cosmological distances. However, for a CS, this quantity is 
$\mathcal{D}\Delta^2/\lambda$, proportional to the lens-observer distance $\mathcal{D}$. 
These fundamental differences have direct implications for their detection and the generation of appropriate GW templates for their identification.

\begin{figure}[t]
\includegraphics[width=0.97\columnwidth]{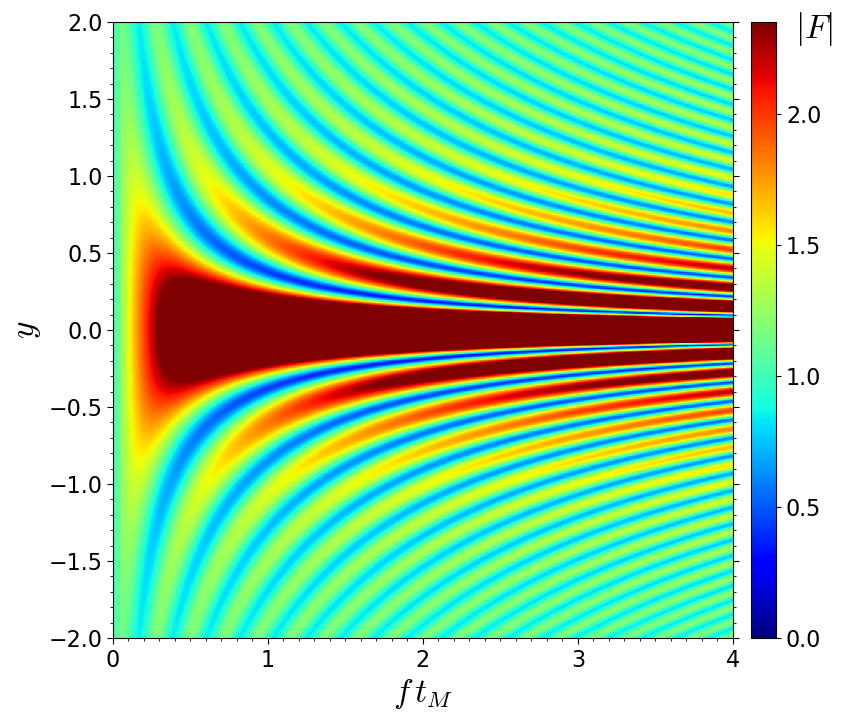}
\caption{Density plot of the transmission factor $|F|$ for a point mass lens, shown as a function of the rescaled frequency $ft_M$ and source position $y = \theta / \theta_{\rm E}$. See Ref.~\cite{BU-JCAP-21} for details.}
\label{fig:F-pml}
\end{figure}

These differences become evident when comparing the absolute value of the transmission factor, $|F|$.  
Since $F$ in both cases depends on two dimensionless quantities, their behavior can be effectively compared using density plots in a two-dimensional space with appropriate scaling, as shown in Fig.~\ref{fig:F-pml} for a PML and Fig.~\ref{fig:F-string-full} for a CS.  
In these plots, the $x$ axis corresponds to the frequency scaled by the characteristic time delay---$t_M$ for a PML [Eq.~\eqref{eq:tM}] and $t_{\Delta}$ for a CS [Eq.~\eqref{eq:t_d}]---while the $y$ axis represents the alignment parameter $y$, defined as the angular position of the source $\theta$ relative to the line of sight, expressed in units of a characteristic angle: the Einstein angle $\theta_{\rm E}$ for a PML, and the parameter $\Delta$ for a CS.

%%%%%%%%%%%%%%%%%%%%%%%%%%%%%%%%%%%%%%%%%%%%%%%%%%%

\section{Cosmic String Lensing Imprints and Detectability}
\label{sec-lensing-cases}

An unlensed GW signal from the source can be described by its frequency-domain strain $\tilde{h}(f)$, obtained via the Fourier transform of the time-domain strain $h(t)$.
When the wave is gravitationally lensed, the observed waveform $\tilde{h}_L(f)$ becomes the product of the lensing transmission factor $F(f)$ and the original unlensed signal,
\beq
    \tilde{h}_L(f) = F(f) \cdot \tilde{h}(f).
\label{eq:lensed_strain}
\eeq

\begin{figure}[t]
\includegraphics[width=\columnwidth]{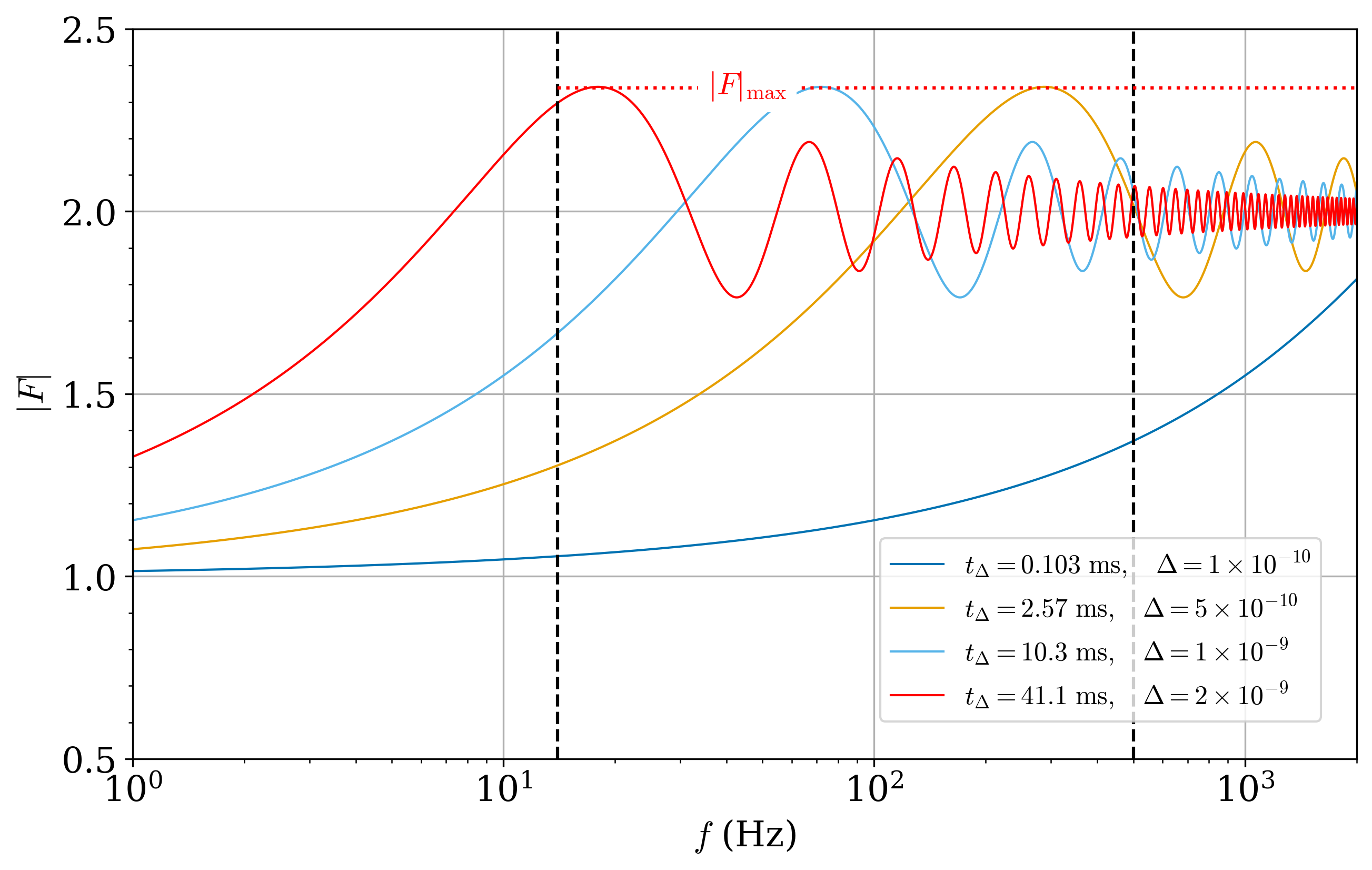}
\caption{Transmission factor $|F|$ as a function of frequency for a source aligned with the line of sight ($y = 0$), showing the effect of varying the string tension $\Delta$ (and the corresponding lensing time delay $t_{\Delta}$).
The LVK detectability band is indicated by the black dashed lines, ranging from $15\,$Hz to $500\,$Hz for illustration.}
\label{fig:F_line-of-sight}
\end{figure}

We are interested in the behavior of the transmission factor within the frequency range accessible to the LVK network, which spans approximately from $f_- \approx 30\,$Hz to $f_+ \approx 1\,$kHz 
\cite{Abbott2020,O4a-Methods}
The transmission factor for cosmic string lensing, given in Eq.~\eqref{eq:F-f-full}, reaches its maximum amplification when the source is perfectly aligned with the line of sight ($y = 0$) occurring at a characteristic frequency of approximately  $f_{00} = (3/4)/t_{\Delta}$, as derived in Appendix \ref{line-of-sight}. 
At this frequency, the magnitude attains its peak value,  $|F|_{\rm max} \approx 2.34$, representing the most prominent feature of wave-optics lensing [see Eq.~\eqref{eq:F_max}]. The amplification consists of a GO contribution factor of 2, corresponding to the two interfering rays, and an additional $\approx 0.34$ arising from the diffracted wave. For the lensing effect to be detectable by the LVK network, the condition $f_{00}<f_+$ must be satisfied. 
If $f_{00}$ significantly exceeds $f_+$, then the amplification rapidly diminishes across all values of the parameter $y$ within the detector band, rendering the lensing signature increasingly difficult to detect.
Figure~\ref{fig:F_line-of-sight} illustrates how the frequency at which the maximum amplification $|F|_{\mathrm{max}}$ occurs shifts to higher values as the string tension $\Delta$, or, equivalently, the lensing time delay $t_{\Delta}$, is decreased.

Enforcing the condition $f_{00} < f_+$ leads to a lower bound on the time delay introduced by the string,
\beq
t_{\Delta} > \frac{3}{4 f_{+}} \approx 0.75\,\mathrm{ms}\,,
\label{eq:lower-bound}
\eeq
for $f_+ = 1\,$kHz.
This requirement, combined with Eq.~\eqref{t_Delta}, leads to a lower bound on the parameter $\Delta$ necessary for the lensing effect to remain detectable, 
\beq
\Delta > 0.85 \times 10^{-8} 
\left(\frac{100\, {\rm Mpc}}{\chi_L} \right)^{1/2} \left(\frac{{\rm Hz}}{f_+}\right)^{1/2},
\label{eq:Delta_max}
\eeq
which, for the LVK frequency range, simplifies to 
\beq
\Delta > 2.7 \times 10^{-10} 
\left(\frac{100\, {\rm Mpc}}{\chi_L} \right)^{1/2}.
\label{eq:Delta_max_LVK}
\eeq
Note that Eq.~\eqref{eq:Delta_max} can also be applied to other types of detectors (e.g., LISA, ET), with $f_+$ representing the corresponding upper frequency bound of the detector's sensitivity band.

\begin{figure}[t]
\includegraphics[width=\columnwidth]{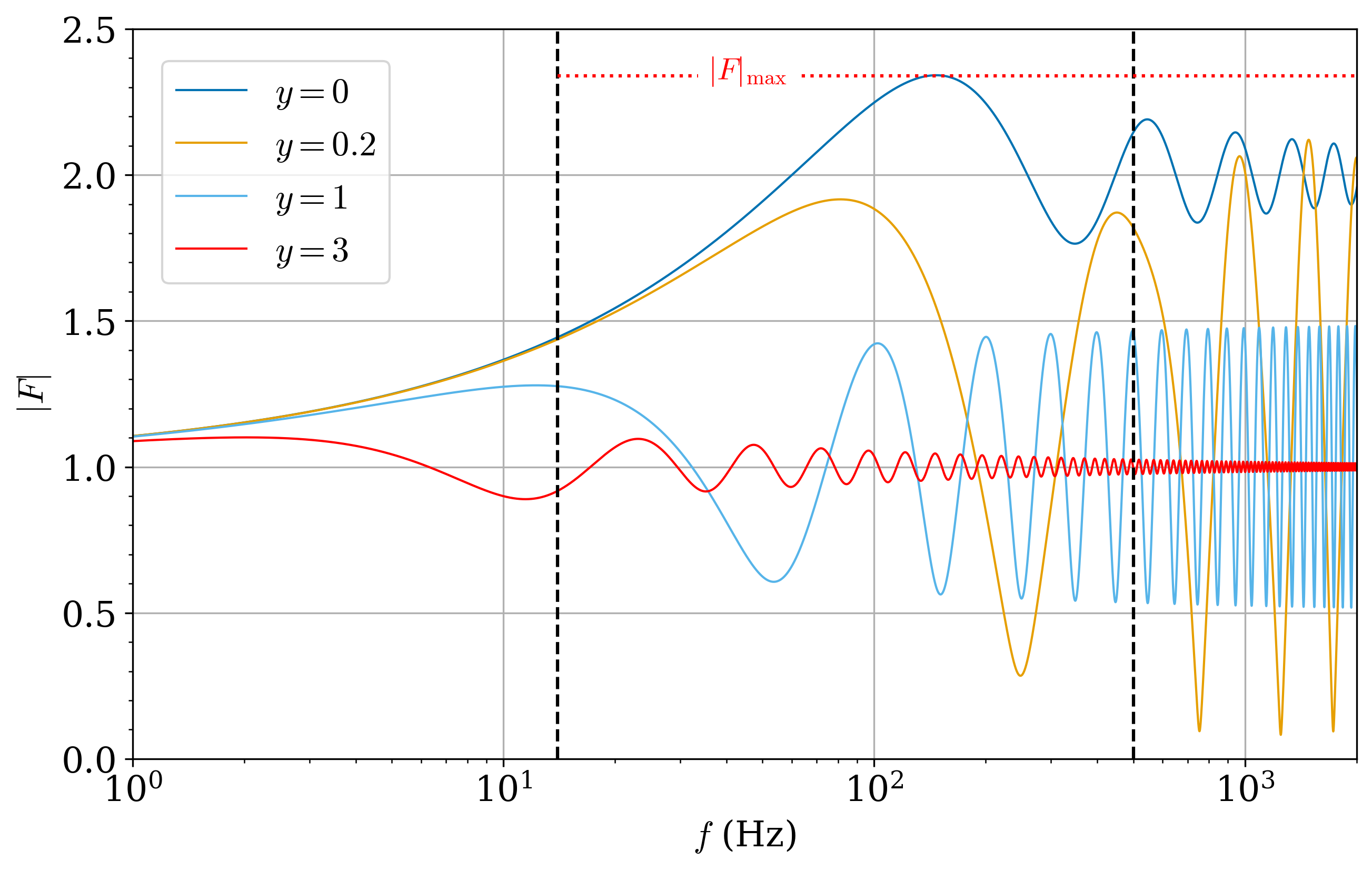}
\caption{Transmission factor $|F|$ as a function of  frequency for fixed time delay $t_{\Delta}=5.04\,$ms, $\Delta=7\times 10^{-10}$, showing the effect of varying the source position $y$ with respect to the line of sight. The LVK detectability band is indicated by the black dashed lines, ranging from $15\,$Hz to $500\,$Hz for illustration.}
\label{fig:F_y-var}
\end{figure}

Suppose the value of $t_{\Delta}$ is such that the maximum amplification $|F|_{\rm max}$ falls within the LVK sensitivity band. As the source position offset $y$ increases from 0 to 3, the transmission factor $|F|$ exhibits damped oscillations as a function of frequency with an overall decreasing trend, but remains above unity, as shown in Fig.~\ref{fig:F_y-var}, indicating that the lensing effect may still be detectable.

\begin{figure}[!htbp]
\includegraphics[width=\columnwidth]{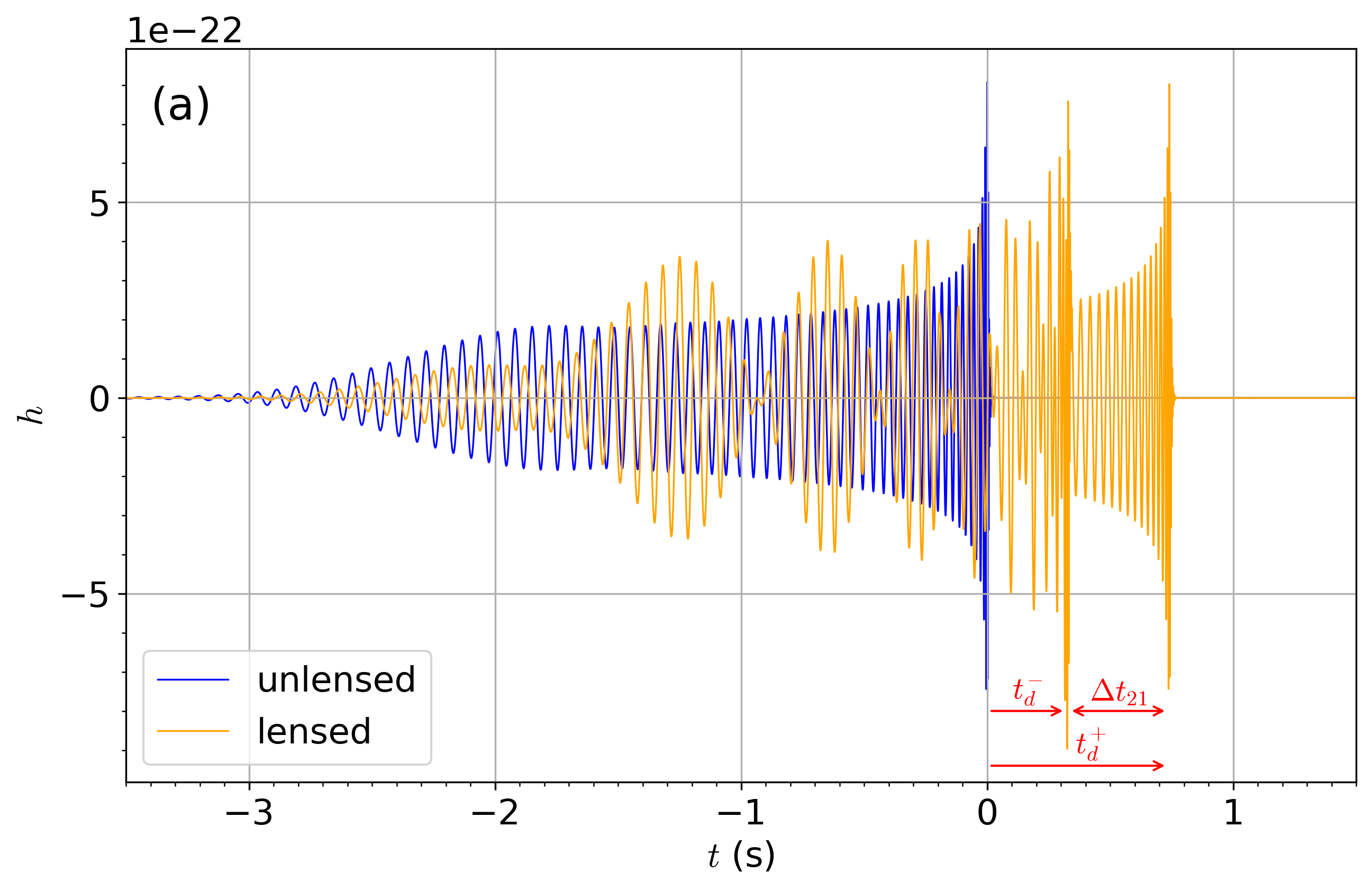}
\includegraphics[width=\columnwidth]{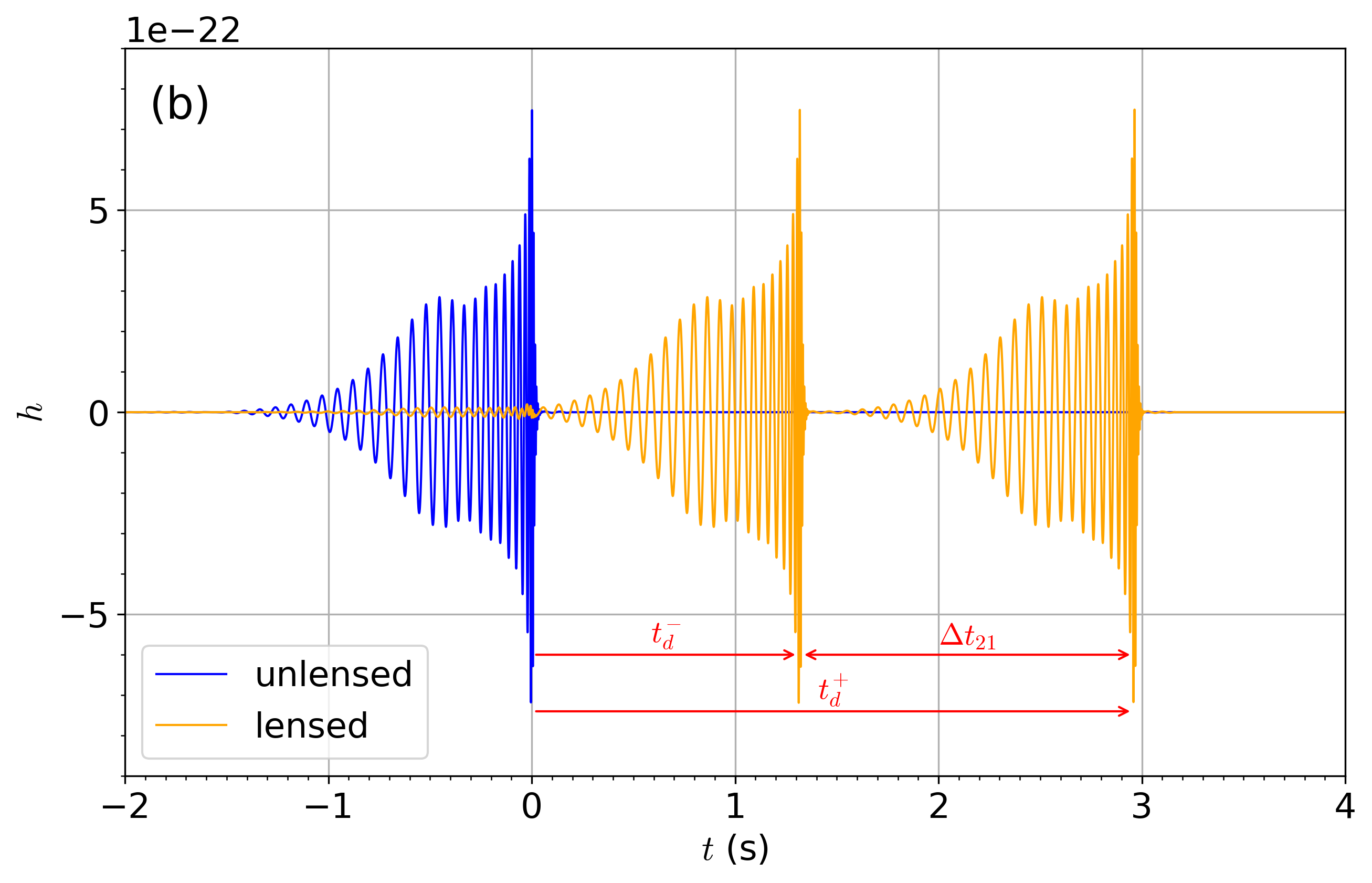}
\caption{
Time-domain strain of a lensed GW (orange) compared to the unlensed waveform (blue).  
Two different lensing regimes are illustrated:  
(a) \emph{microlensing} regime, where the two images overlap and interfere, producing a beating pattern, and   
(b) \emph{strong lensing} regime, with two images separated in time.  
The time delays are indicated by red lines and are reported in Table~\ref{tab:lensing_params}.
}
\label{fig:h-t}
\end{figure}

To better understand the gravitational lensing effect, it is useful to visualize how it imprints on the gravitational waveform. The time-domain strain $h(t)$ of the lensed signal is obtained by taking the inverse Fourier transform of the frequency-domain waveform $\tilde{h}_L(f)$ modified by the transmission factor, as given in Eq.~\eqref{eq:lensed_strain}. The resulting lensed waveform is shown in Fig.~\ref{fig:h-t}. Correspondingly, Table \ref{tab:lensing_params} summarizes the values of the parameters, both for the BBH source and the lens, used to generate the figure.
Here, we distinguish between two qualitatively different lensing regimes:
(i) {\em microlensing} regime---when the time delay between the two images is small, the waveforms overlap and interfere, producing a characteristic beating pattern, and 
(ii) {\em strong lensing} regime---when the time delay is large enough for the two lensed images to be well separated in time.
In the microlensing case [Fig.~\ref{fig:h-t}(a)], the earliest signal consists of a beating pattern from the interference of both images, followed by the first merger peak. The second image arrives later without interference, keeping a shape identical to the original waveform.
In the strong lensing case  [Fig.~\ref{fig:h-t}(b)], both images are exact time-shifted replicas of the original signal---a direct consequence of the topological nature of the lensing.
The presence of two distinct merger peaks in both regimes is a unique signature of gravitational lensing, distinguishing it from other effects that can produce beating.

The arrival times of the first and second images, measured relative to the unlensed signal, can be estimated based on Eqs.~\eqref{eq:F-f-full} (full wave) or \eqref{eq:F-GTD} (GO), both of which contain phase delays of the form $\exp(-2\pi \ii f  t_d^\pm)$. 
These phase terms originate from the full-wave solution~\eqref{eq:u-diff}, expressed as a sum of Sommerfeld's edge-diffracted waves with geometrically defined phases $kr \cos(\Delta \pm \theta)$, corresponding to optical path differences. 
This formulation directly yields the time delay between the two images as
\begin{equation}
\Delta t_{21} = t_d^+ - t_d^- = 2y\,t_{\Delta},
\label{eq:t21}
\end{equation}
which has the same form as Eq.~\eqref{eq:t21-PML} for the PML.
For the source and lens parameters listed in Table~\ref{tab:lensing_params}, the time delays obtained by modulating the waveform with $F(f)$ and applying an inverse Fourier transform show excellent agreement with the analytical predictions from Eq.~\eqref{eq:t21}.

% PARAMETERS AND TIME DELAYS TABLE
\begin{table}[t]
\caption{Source and lens parameters (separated by horizontal lines), along with predicted time delays for the two lensing cases shown in Fig.~\ref{fig:h-t}.
The waveforms were generated using the \texttt{IMRPhenomXPHM} approximant~\cite{pratten2021IMRPhenomXPHM}, with a lower frequency cutoff of $f_{\text{low}} = 14$ Hz.
}
\label{tab:lensing_params}
\centering
\begin{tabular}{lcc}
\hline\hline
Parameter & (i) Microlensing & (ii) Strong lensing \\
\hline
$m_1$, $m_2$ ($M_\odot$) & 36, 29 & 60, 60 \\
Distance (Mpc) & 1000 & 2000 \\
\hline
$y$ & 0.2 & 0.2 \\
$\Delta$ & $10^{-8}$ & $10^{-8}$ \\
$\chi_L$ (Mpc) & 100 & 400 \\
\hline
%Time scale 
$t_\Delta$ (s) & 1.0287 & 4.1147 \\
$\Delta t_{21}$ (s) [Eq.~\eqref{eq:t21}] & 0.4115 & 1.6459 \\
%Time delay numerical 
$\Delta t_{21}$ (s) [Fig.~\ref{fig:h-t}] & 0.4148 & 1.6458 \\
\hline\hline
\end{tabular}
\end{table}

%%%%%%%%%%%%%%%%%%%%%%%%%%%%%%%%%%%%%%%%%%%%%%%%%%%

\section{Mismodeling biases and identification of CS signatures.}
\label{sec-selection-bias}

For the detection of compact binary coalescence signals, most GW search pipelines rely on matched filtering~\cite{abbott2020signal_processing_guide,O4a-Methods}. However, the absence of lensed waveforms in the template banks used in matched filtering introduces a systematic bias. In this work, we investigate the impact of this bias in the case of a CS lens.
We disentangle two sources of detection bias when analyzing lensed signals with unlensed template banks. First, we quantify the loss of matched-filter signal-to-noise ratio (SNR) arising from the waveform mismatch between the true lensed signal and the unlensed templates. Second, we assess the additional reduction in detection significance caused by the $\chi^2$ signal-consistency test~\cite{allen2005chisq}, which is employed in the {\sc PyCBC} pipeline~\cite{dalcanton2014pycbc, usman2016pycbc, nitz2017pycbc} for glitch mitigation.  
Reference \cite{chan-2024-Detectability_lensing_matched_filter} explicitly accounts for this effect via full injection campaigns of PML-lensed signals analyzed with unlensed templates.
Furthermore, we investigate the extent to which CS-lensed signals can be distinguished from those lensed by a PML and from unlensed signals by computing the odds ratio. To do this we estimate the prior ratio from detection rates and compute the evidence ratio from parameter estimations using the {\sc Bilby} framework.

\subsection{Matched filtering and consistency tests}

A matched filtering algorithm typically searches for a transient signal by computing the correlation between the detector data and a bank of template waveforms.  This correlation is given by \cite{maggiore-07,allen2012findchirp}
\begin{equation}
    \label{eq:cmplx_snr_ts}
    z_j(t)=4\int_0^{\infty}\frac{d(f)h_j^*(f)}{S_n(f)}e^{2\pi \ii ft} \dd f\,,
\end{equation}
where $d(f)$ is the data employed in the search, $h_j(f)$ is the $j$th template in the bank, and $S_n(f)$ is the detector's noise power spectral density (PSD). 
The matched-filter output $z_j(t)$  is a complex time series, and its magnitude $\rho(t)=|z_j(t)|$, defines the SNR. A candidate event is identified when $\rho(t)$ exceeds a predefined threshold. Under the assumption of stationary Gaussian noise, the SNR is maximized when the template $h(f)$ matches the true signal~\cite{abbott2020signal_processing_guide}. 

The value for the SNR is influenced by the similarity between the actual signal buried in the noise, denoted as $s(t)$, and the templates used in the matched filtering process. This similarity can be quantified using the  \textit{mismatch}, defined as 
\begin{equation}
    \label{eq:mismatch}
    M_j = 1-\argmax_{\phi, t_c} \frac{( s|h_j)}{\sqrt{( s|s)( h_j|h_j)}}\,,
\end{equation}
where $\phi$ denotes the signal phase, $t_c$ is the coalescence time, and the subscript $j$ refers to the $j$th waveform in the template bank. A mismatch of zero corresponds to perfect agreement between the signal and the template, yielding the optimal SNR, $\rho_{\rm opt}$, which serves as an upper bound:
\begin{equation}
 \rho_{\rm opt}^2 = 4 \int_{f_{\rm min}}^{f_{\rm max}} \frac{|\tilde{h}(f)|^2}{S_n(f)} \dd f\,,
\end{equation}
where $f_{\rm min}$ and $f_{\rm max}$ are the minimum and maximum frequencies under consideration from our sampling choice. As the mismatch increases, the achievable SNR decreases accordingly.

Signal consistency tests, such as the $\chi^2$ test, are traditionally used to suppress non-astrophysical transients, or glitches, that may mimic real signals \cite{davis2022glitches_review}. However, their utility extends beyond glitch rejection. These tests also evaluate whether the signal's contribution to the matched-filter SNR is consistent across frequency bands, helping to distinguish true signals from structured noise \cite{allen2005chisq}. Additionally, they provide a measure of how well the data matches the expected waveform, which can be used to improve event ranking, assess signal fidelity, and support multidetector coincidence analyses \cite{reweighted_SNR, usman2016pycbc}.

For our assessment, we employ the {\sc PyCBC} pipeline. This pipeline computes a reweighted SNR time series $\hat{\rho}$ that downranks signals with unfavorable $\chi^2$ test results as follows \cite{reweighted_SNR, usman2016pycbc}:
\begin{equation}
    \label{eq:pycbc_snr_rewighting}
    \hat{\rho} = \rho \times
    \begin{cases} 
        1, & \chi^2 \leq \nu\,, \\
        \left[ \frac{1}{2} + \frac{1}{2} \left(\chi^2/\nu\right)^3 \right]^{-1/6}, & \chi^2 > \nu\,,
    \end{cases}
\end{equation}
where $\nu$ is the number of degrees of freedom in the $\chi^2$ test.
Although this strategy is effective for suppressing non-astrophysical transients, it may also downrank genuine signals whose frequency content deviates from the templates used. In our case, the transmission factor in Eq.~\eqref{eq:F-f-full} introduces a frequency modulation that differs from the expectations of the $\chi^2$ test when matched filtering is performed using standard (unlensed) templates. This can lead to an increased mismatch and a reduced SNR due to: (i) a lower peak in the SNR time series, and (ii) 
additional penalization from the signal consistency test.

\begin{figure*}[t]
    \centering
    \includegraphics[width=0.3\linewidth]{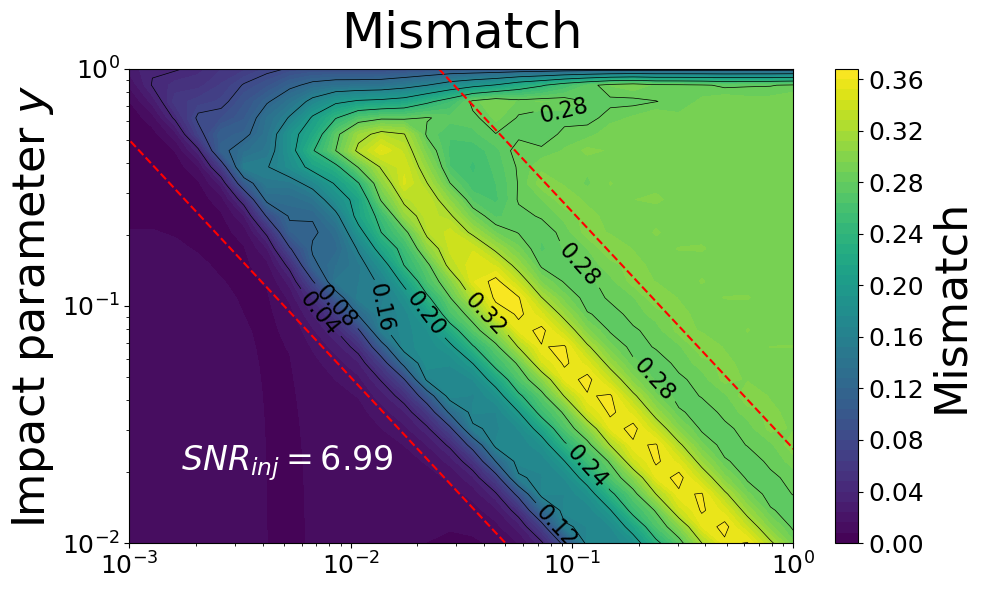}
    \includegraphics[width=0.3\linewidth]{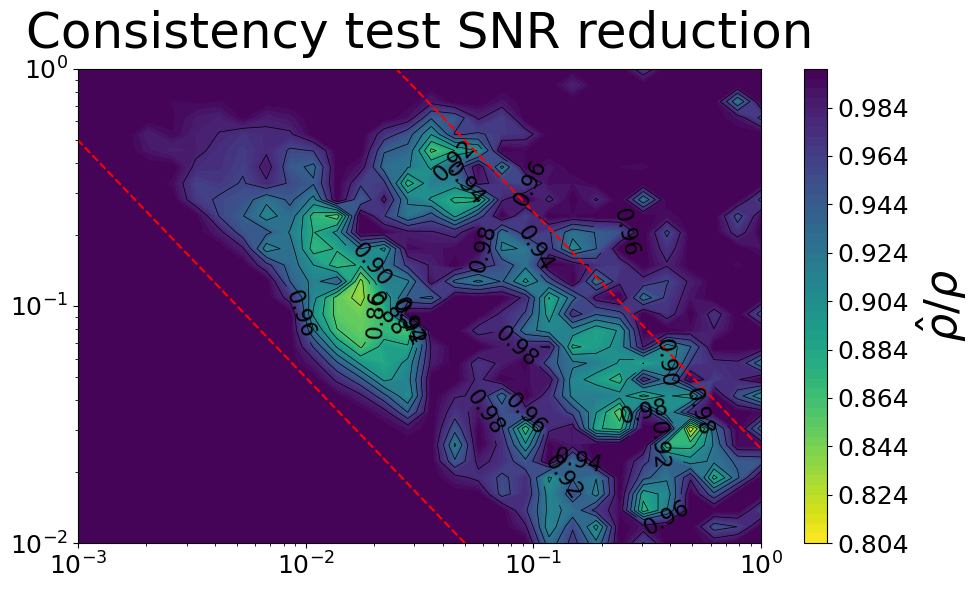}
    \includegraphics[width=0.3\linewidth]{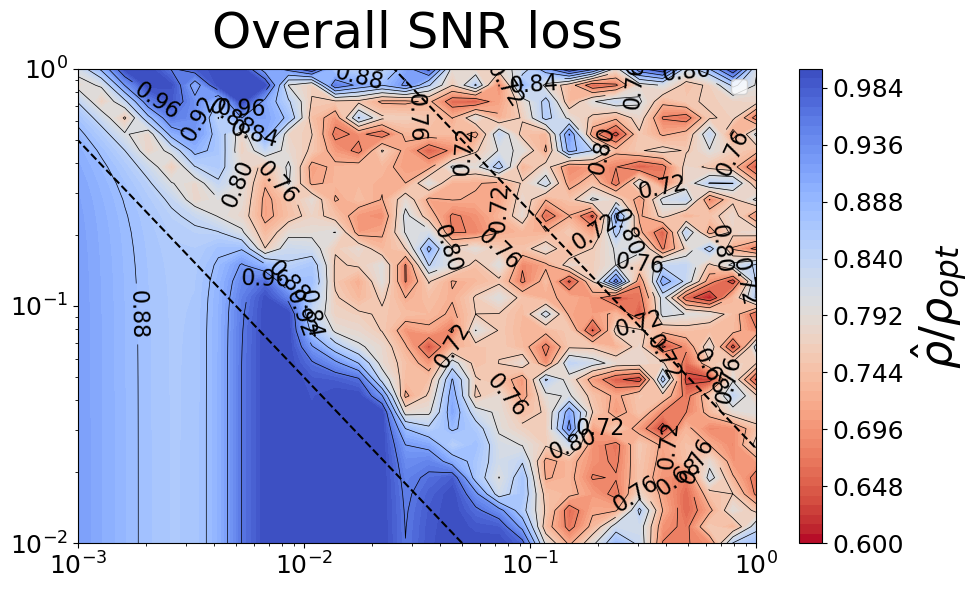}\\
    \includegraphics[width=0.3\linewidth]{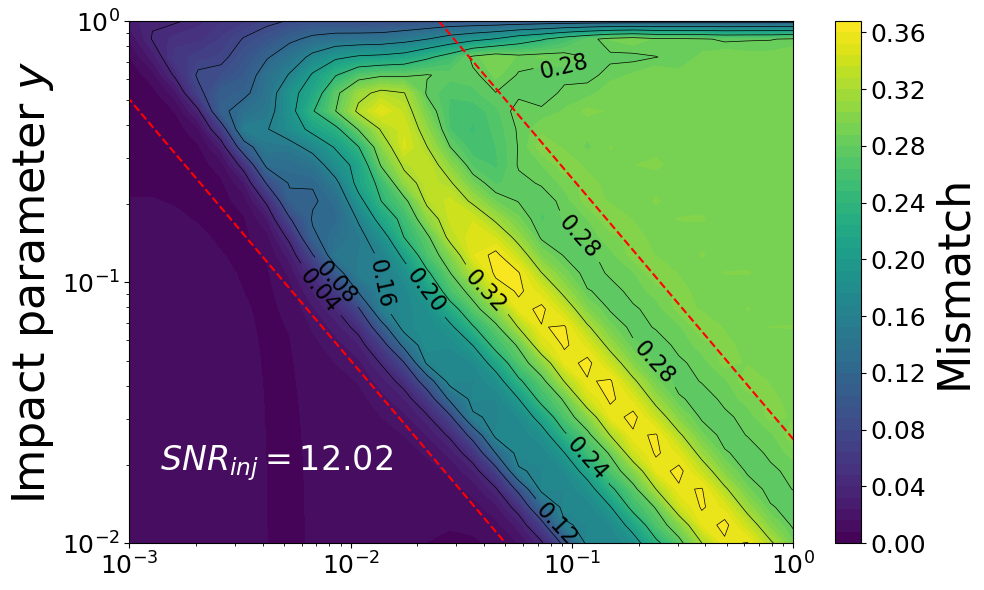}
    \includegraphics[width=0.3\linewidth]{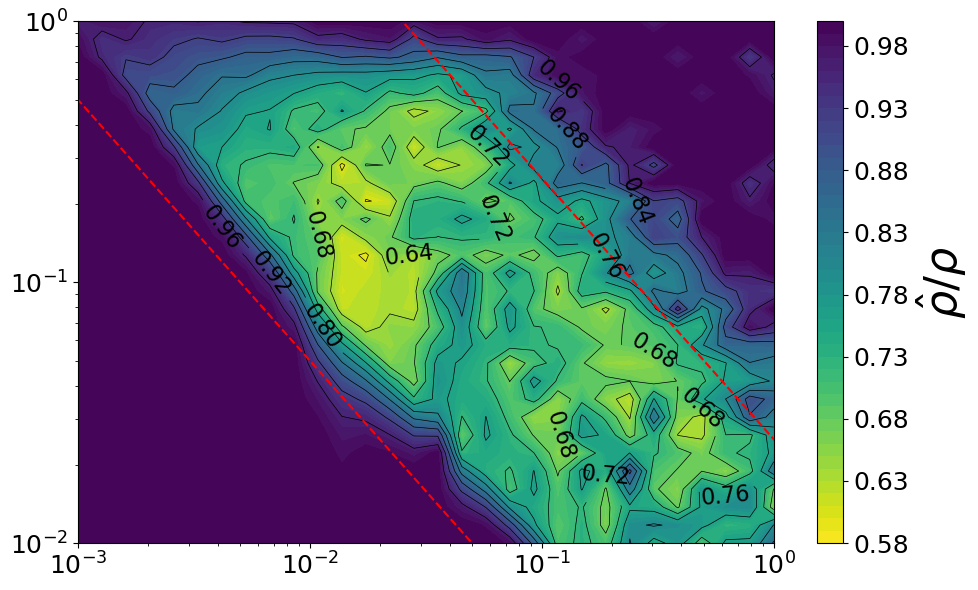}
    \includegraphics[width=0.3\linewidth]{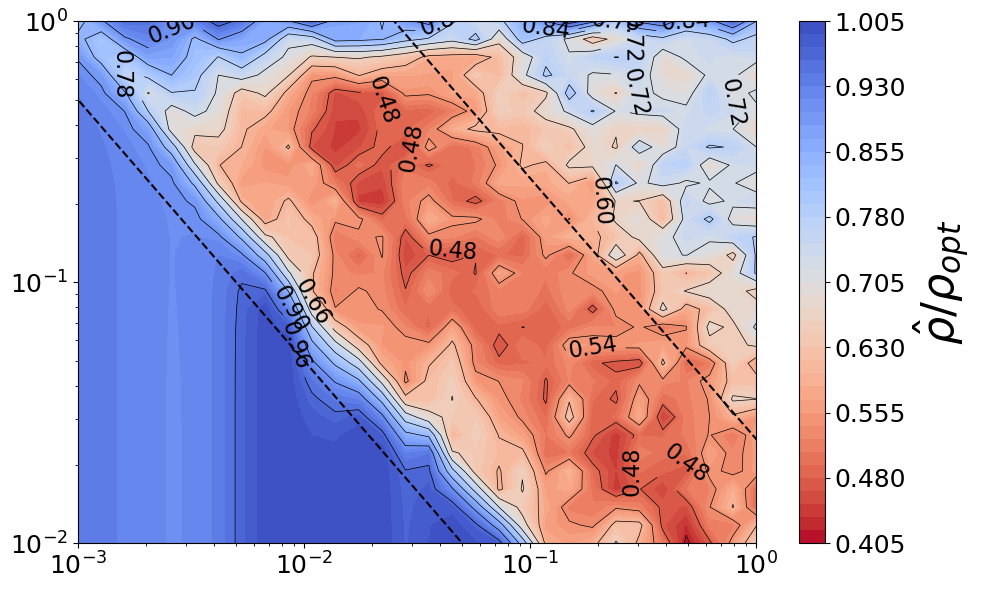}\\
    \includegraphics[width=0.3\linewidth]{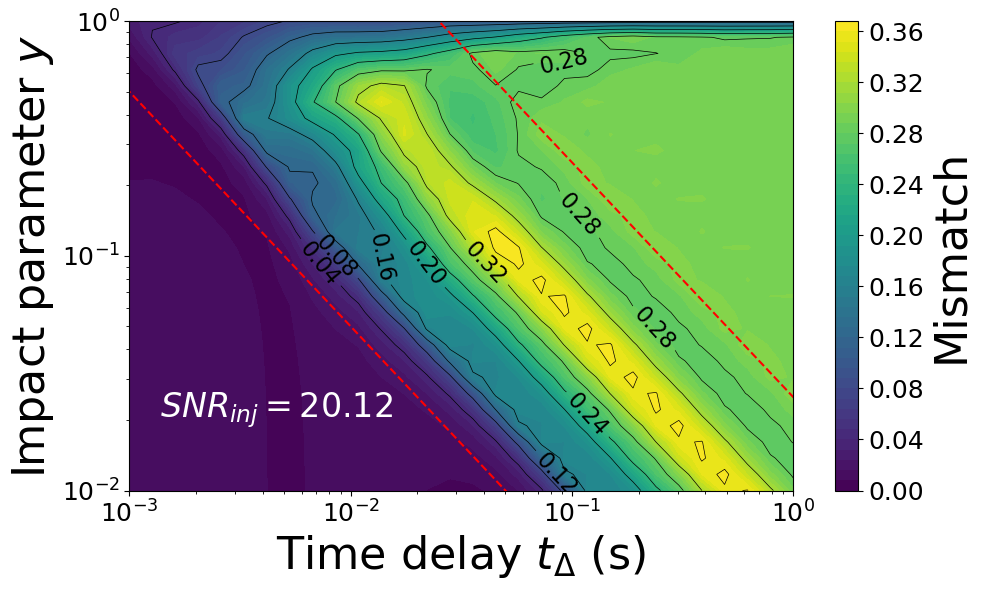}
    \includegraphics[width=0.3\linewidth]{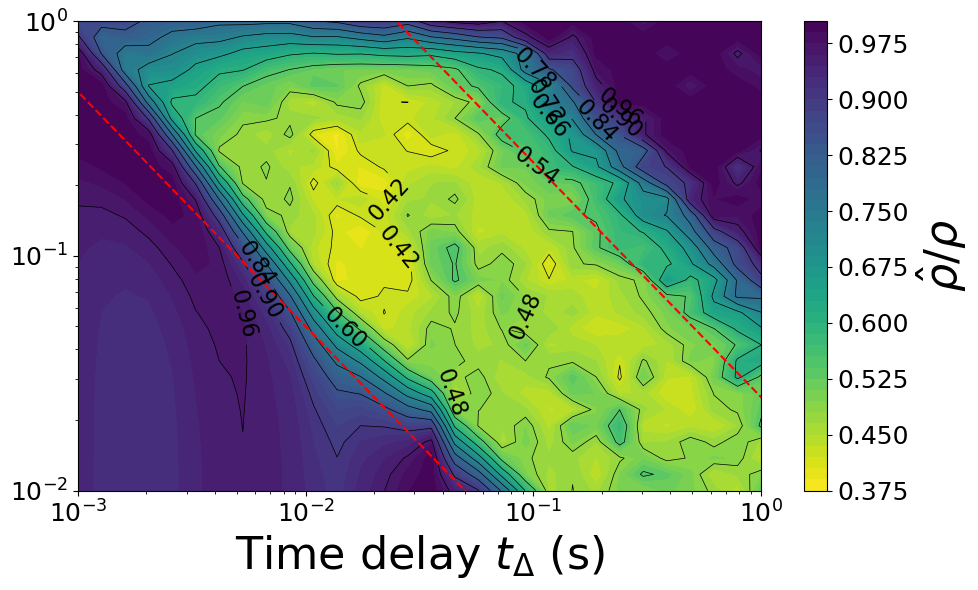}
    \includegraphics[width=0.3\linewidth]{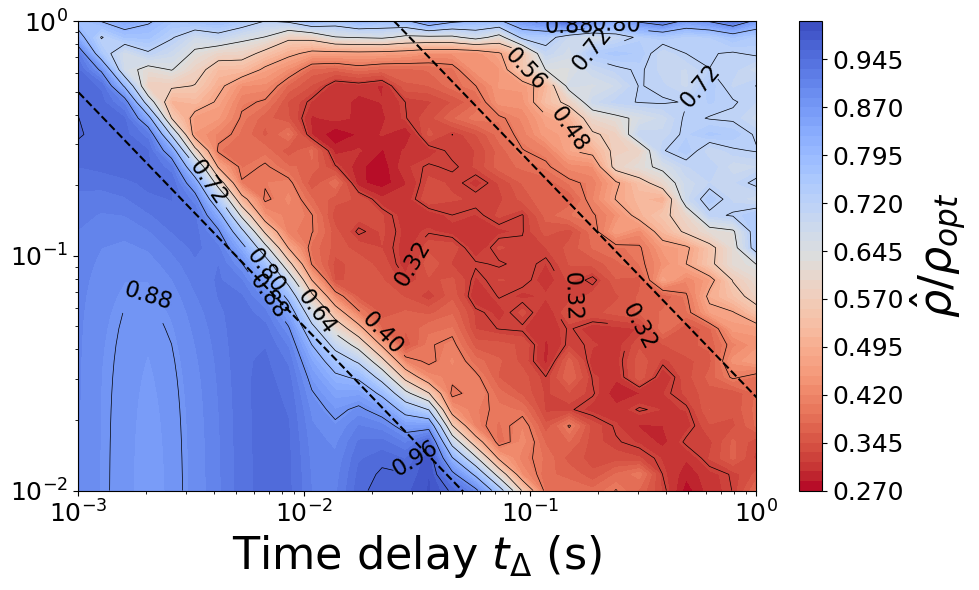}
    \caption{Impact of template mismatch and signal consistency tests on the SNR of cosmic string-lensed signals. Each row corresponds to one of the injections listed in Table~\ref{tab:waveform_parameters}, ordered by increasing injected SNR from top to bottom. The columns show, from left to right: (i) the mismatch between lensed and unlensed waveforms, (ii) the SNR loss due to the $\chi^2$ consistency test, quantified by the ratio $\hat{\rho}/\rho$, and (iii) the total SNR loss relative to the optimal injected value, given by $\hat{\rho}/\rho_{\rm opt}$. 
    Dashed red lines (black in the rightmost column) delimit the region in the $y$–$t_\Delta$ parameter space for which $f_\Delta$ lies within the detectors' sensitivity band, $20\,\mathrm{Hz} < f_\Delta < 1000\,\mathrm{Hz}$, where $f_\Delta$ is defined in Eq.~\eqref{eq:f_delta}.}
\label{fig:snr_effects}
\end{figure*}

\subsection{Quantitative analysis of the lensing mismodeling bias}

Using the {\sc PyCBC} package, we inject three BBH signals generated with the {\sc IMRPhenomXPHM}~\cite{pratten2021IMRPhenomXPHM} waveform,
produced with the default multi-scale analysis spin prescription 
into simulated noise based on the Advanced LIGO PSD under the high-sensitivity settings of the T2000012-v2 technical document\footnote{\href{https://dcc.ligo.org/LIGO-T2000012/public}{https://dcc.ligo.org/LIGO-T2000012/public}}. The parameters of our BBH waveforms are listed in Table~\ref{tab:waveform_parameters}. We then apply the CS transmission factor \eqref{eq:F-f-full}, exploring lensing parameters in the ranges $t_\Delta\in[10^{-3},1]\,$s and $y\in[10^{-2},1]$, over a $30\times 30$ equally spaced grid. This range of parameters, together with the chosen signal duration, has been selected to ensure that the lensing effects remain in the microlensing regime [Fig.\ref{fig:h-t}(a)]. In the strong-lensing regime, the original waveform is not significantly modulated; instead, two time-delayed copies of the signal are produced [Fig.\ref{fig:h-t}(b)]. In this situation, the mismatch between the unlensed and lensed waveforms is null, so no mismodeling bias arises. For each lensed signal, we compute the mismatch with the corresponding unlensed waveform, the matched-filter SNR time series, the $\chi^2$ test, and the reweighted SNR. The results are linearly interpolated to produce a 2D visualization across the full range of lensing parameters.
\begin{table}[b!]
\caption{BBH signal parameters of the three injections used in this study. $\mathcal{M}$ is the chirp mass, $q$ is the mass ratio, and $M_{\rm T}$ is the total mass of the binary. $\chi_{iz}$ denotes the $z$ component of the dimensionless spin parameter for the $i$th black hole (the other components are negligible). $D_L$ is the luminosity distance to the source, $\varphi_c$ is the phase at coalescence, while $\alpha$ and $\delta$ are the right ascension and declination angles, respectively. The injection SNR refers to the SNR obtained for the unlensed injection.}
\label{tab:waveform_parameters}
\vspace{2mm}
\centering
\begin{tabular}{c|c}
\hline
\hline
$\mathcal{M}$ & 28.095 $\rm M_\odot$ \\\hline
$q=M_2/M_1$ & 0.805 \\\hline
$M_T$ & 65 $\rm M_\odot$ \\\hline
$\chi_{1z}$ & 0.4 \\\hline
$\chi_{2z}$ & 0.3 \\\hline
$D_L$ & [6205.03, 3608.04, 2156.78] Mpc \\\hline
$\varphi_c$ & 1.3 rad \\\hline
$\alpha$ & 1.375 rad \\\hline
$\delta$ & $-1.2108$ rad \\\hline
Injection SNR & [6.99, 12.02, 20.12]\\\hline \hline
\end{tabular}
\end{table}

Figure~\ref{fig:snr_effects} summarizes our findings.  The left column shows the mismatch between lensed and unlensed waveforms as a function of impact parameter $y$ and time delay $t_\Delta$. 
The central column displays the SNR loss due to the consistency test, quantified by the ratio $\hat{\rho}/\rho$, where $\hat{\rho}$ is the reweighted SNR from Eq.~\eqref{eq:pycbc_snr_rewighting}, and $\rho$  is the matched-filter SNR obtained using an unlensed template bank. 
The right column shows the total SNR loss in the matched-filtering process, given by the ratio $\hat{\rho}/\rho_{\rm opt}$, where $\rho_{\rm opt}$  is the optimal SNR of the injected lensed signal. Each row in the figure corresponds to one of the three injections listed in Table~\ref{tab:waveform_parameters}, ordered by increasing SNR from top to bottom. 
To relate these results to the sensitivity band of the detectors, the relevant frequency range is indicated in Fig.~\ref{fig:snr_effects} (see the figure caption for details).
%To relate these results to the frequency of the beating pattern $f_\Delta$ (defined in Eq.~\eqref{eq:f_delta}), combinations of $y$ and $t_\Delta$ that give frequencies between 20 Hz and 1000 Hz are indicated by dashed red lines (black in the rightmost column). 

By inspecting Eqs.~\eqref{eq:F-GTD} and \eqref{eq:f_delta}, one can define a critical value for the characteristic time delay, $t_{\Delta}$, at which the GO term begins to dominate over the diffraction term. We denote this critical value as $t_{\Delta}^{\mathrm{GO}}$. Using the time evolution of the GW frequency during the (adiabatic) inspiral phase of a compact binary coalescence \cite{maggiore-07},
\begin{equation}
f(t) = \frac{1}{8\pi}\left(\frac{c^3}{G\mathcal{M}}\right)^{5/8}\left[\frac{5}{t_{\rm coal}-t}\right]^{3/8},
\end{equation}
where $t_{\rm coal}$ represents the time of coalescence, a reference frequency depends only on the chirp mass,
\begin{equation}
%f_{\mathrm{GO}} = \frac{1}{8\pi}\left(\frac{c^3}
f^* = \frac{1}{8\pi}\left(\frac{c^3}
{G\mathcal{M}}\right)^{5/8}\,.
\end{equation}
When the characteristic frequency $f_\Delta$, given by Eq.~\eqref{eq:f_delta}, falls below $f^*$, the diffraction effects become negligible. This defines a critical value for $f_\Delta$,
\begin{equation}
    f_{\Delta}^{\mathrm{GO}}=\frac{1}{8\pi}\left(\frac{c^3}{G\mathcal{M}}\right)^{5/8}=\frac{1}{2t_{\Delta}^{\mathrm{GO}}y},
\end{equation}
which, in turn, defines the critical GO time delay mentioned above,
\begin{equation}
t_\Delta^{\mathrm{GO}} = \frac{4\pi}{y}\left(\frac{G}{c^3}\right)^{5/8}\mathcal{M}^{5/8}
\;\approx \; \frac{0.012\,{\mathrm s}}{y}\,\left(\frac{\mathcal{M}}{M_{\odot}}\right)^{5/8}.
\end{equation}
Notice the dependence on $y$, which states that when observer, string and source are aligned, the diffraction effects are too significant to be ignored. This regime is visible in the central column of Fig.~\ref{fig:snr_effects}, in the upper-right corner of the plots, where $t_\Delta$ and $y$ lie within the GO regime. In this case, the transmission factor contributions are more uniform across frequency, leading to more constant SNR contributions and effectively reducing the penalty from the consistency test.

We also observe how the mismatch between the template bank and the detected signal impacts the SNR at different stages of the analysis. There are two distinct contributions to the overall SNR reduction: one arising from the waveform mismatch, which directly affects the matched-filter output and another from the $\chi^2$ test. 
Interestingly enough, although both effects stem from the mismatch between the template and the signal, the $\chi^2$ test introduces a distinct pattern in the distribution of SNR downranking. Moreover, the penalization from the $\chi^2$ test increases with the SNR, in contrast to the mismatch-induced reduction, which is largely independent of the signal strength.
The most significant SNR losses occur in the region of lensing parameters where the beating frequency $f_\Delta$ lies within the range $[20, 1000]\,$Hz, which overlaps well with the sensitivity band of the LVK detectors. In extreme cases, the total SNR reduction can reach up to $\sim70\%$ of the injected SNR, with $\sim60\%$ attributable to the $\chi^2$ consistency test alone. However, this reduction is less pronounced for injections with lower intrinsic SNR. 

The observation that the mismatch peaks at intermediate frequencies---where the detectors are most sensitive---can be understood as a consequence of the frequency-dependent weighting in the matched-filtering process. Since the detector's sensitivity is highest in this band, even small discrepancies between the lensed and unlensed waveforms in this region contribute significantly to the mismatch integral. This effect is not necessarily due to the waveform model itself, but rather to the fact that the lensing-induced modulation alters the signal most noticeably in the frequency range where the detector is most responsive.
We expect this behavior to be qualitatively similar across different waveform approximants, as the mismatch is dominated by the modulation introduced by the transmission factor rather than the intrinsic waveform morphology. However, a detailed comparison using alternative approximants (e.g., {\sc SEOBNRv4PHM}) could help quantify the robustness of this feature.

\begin{figure*}[t!]
    \centering
    \includegraphics[width=0.3\linewidth]{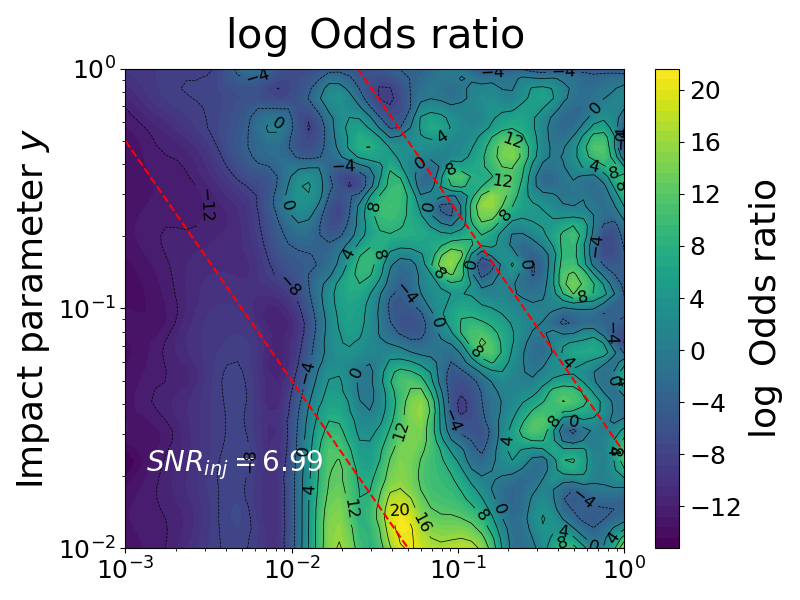}
    \includegraphics[width=0.3\linewidth]{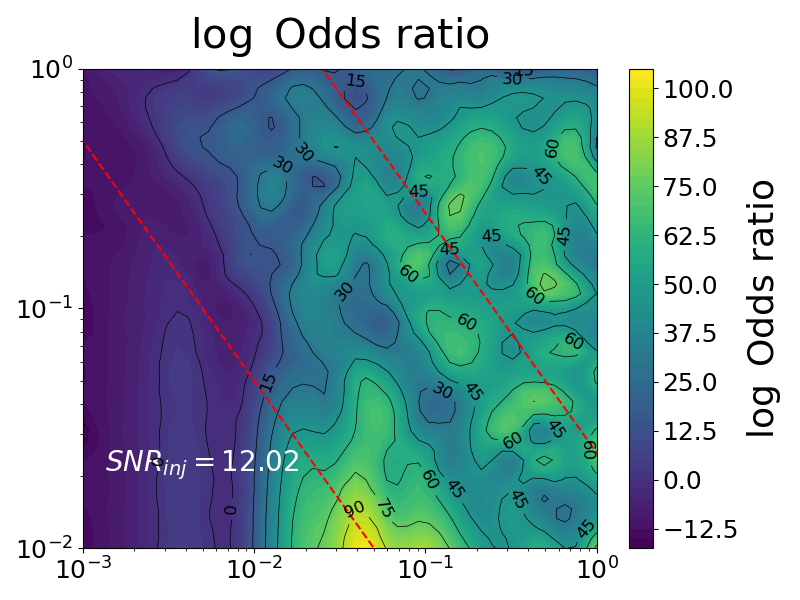}
    \includegraphics[width=0.3\linewidth]{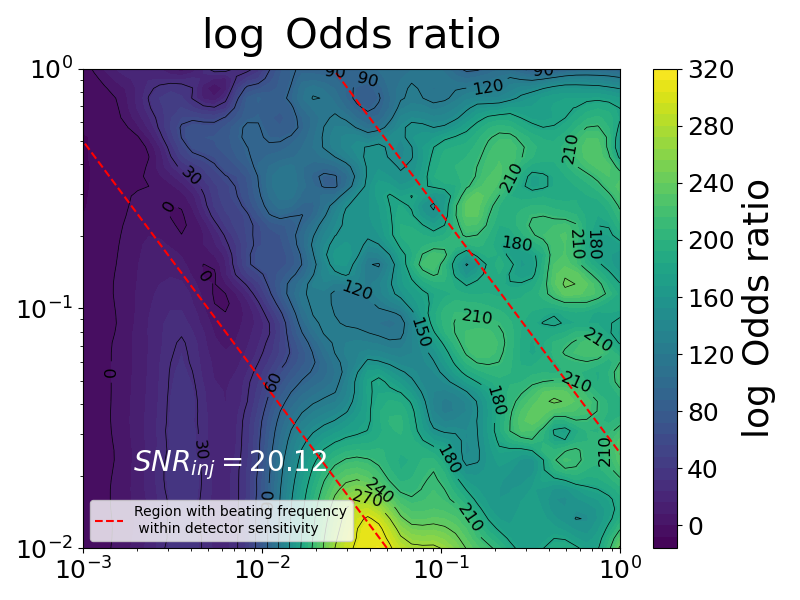}\\
    \includegraphics[width=0.3\linewidth]{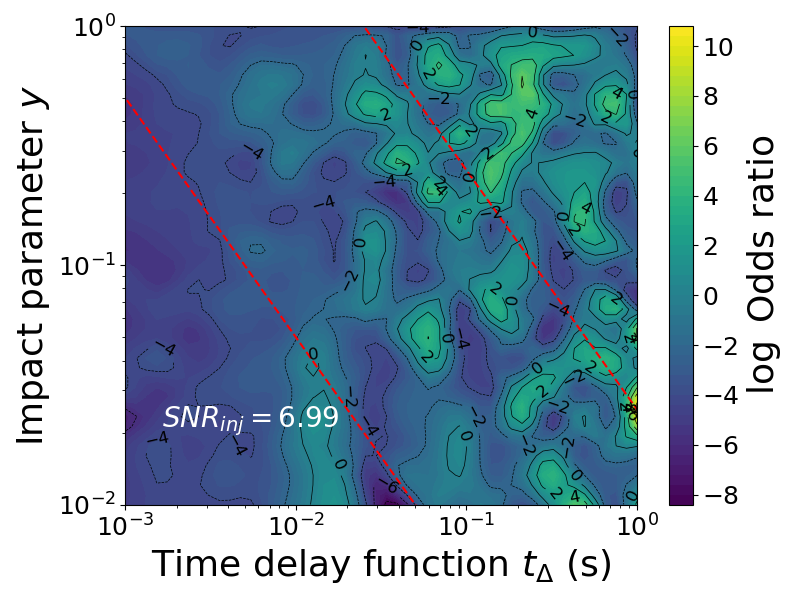}
    \includegraphics[width=0.3\linewidth]{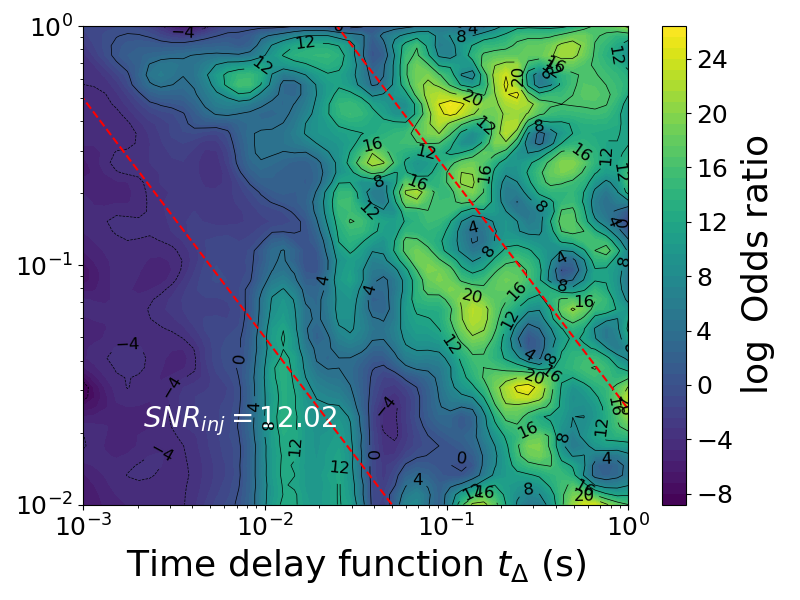}
    \includegraphics[width=0.3\linewidth]{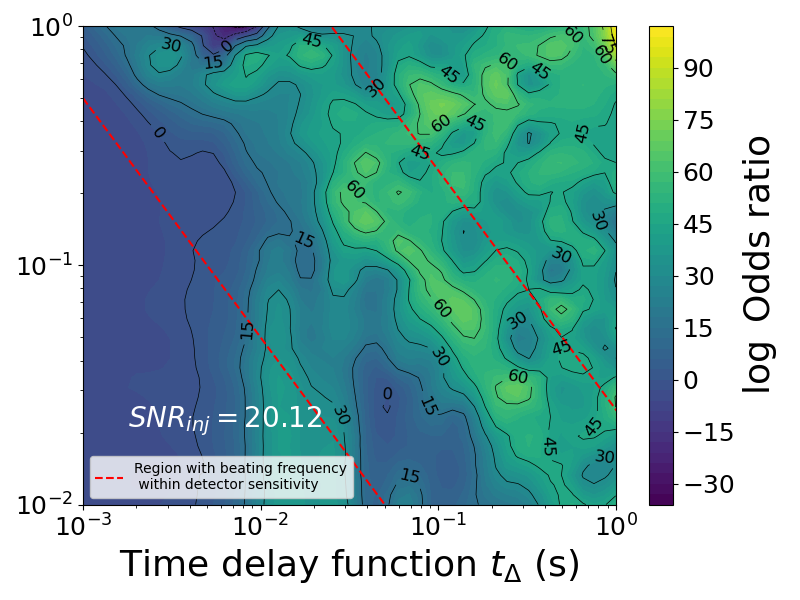}
    \caption{Log odds ratio distributions as a function of lensing parameters: the time delay $ t_\Delta$ and impact parameter $y$. The top row compares unlensed models with signals lensed by CSs. Positive values of the log odds ratio favor the CS lensing hypothesis, while negative values favor the unlensed model. 
    The bottom row presents the same analysis, comparing CS lensing against lensing by a PML. As in Fig.~\ref{fig:snr_effects}, dashed red lines (black in the rightmost column) delimit the region in the $y$–$t_\Delta$ parameter space for which $f_\Delta$ lies within the detectors' sensitivity band, $20\,\mathrm{Hz} < f_\Delta < 1000\,\mathrm{Hz}$.} 
    \label{fig:detectability}
\end{figure*}

\subsection{Distinguishability of Lensed Signals}

The mismatch between the signals in the template bank and the detected signal introduces a bias in matched filtering-based searches. However, this mismatch also provides an opportunity to distinguish between different waveform models. To explore this, we employ Bayesian parameter estimation, which offers a framework for comparing the probabilities of competing hypotheses given a segment of strain data. The ratio of these probabilities is known as the \textit{odds ratio}:
\begin{equation}\label{eq:odds_ratio}
    \frac{p(\mathcal{H}_1|d)}{p(\mathcal{H}_2|d)} = \frac{p(d|\mathcal{H}_1)}{p(d|\mathcal{H}_2)}\times\frac{p(\mathcal{H}_1)}{p(\mathcal{H}_2)}\,,
\end{equation}
where $\mathcal{H}_i$ represents the hypothesis under exploration, which can be either having a waveform lensed by a CS, by a PML or unlensed, $d$ refers to the observed data and $p$ represents the corresponding conditional probability. The first term on the right-hand side is the \textit{Bayes factor} (BF), which quantifies the ratio of evidences for the two hypotheses. It is typically computed using sampling methods that estimate the posterior distribution $p(\theta|d, \mathcal{H}_i)$, allowing the evidence to be evaluated by marginalizing over the model parameters $\theta$: \begin{equation}\label{eq:bayer_factor_evidence}
    p(d|\mathcal{H}_i) = \int p(d|\mathcal{H}_i, \theta)p(\theta|\mathcal{H}_i)\dd\theta\,.
\end{equation}
In our analysis, we used the {\sc Bilby}~\cite{bilby_paper} library and the {\sc nessai}~\cite{nessai} sampler for the BF calculation. These libraries implement algorithms for estimating the posterior probability distribution $p(\theta|d)$ using nested sampling. The inference relies on a probabilistic description of the detector noise, typically assumed to be Gaussian. This approximation is accurate for sufficiently short, glitch-free signals, for which non-stationarity in the noise can be neglected. In our case, the noise is generated directly from the detector sensitivity curve and is therefore exactly Gaussian by construction.

Given this noise model, we can define a likelihood function $\mathcal{L}(d|\theta)$. Combined with specified prior distributions on the waveform parameters, the likelihood enables computation of the evidence (or marginal likelihood) by integrating the likelihood over the prior volume. Nested sampling performs this calculation by drawing parameter samples from the prior and iteratively replacing the lowest-likelihood sample with a new one of higher likelihood. This procedure transforms the multidimensional evidence integral into a one-dimensional integral over the decreasing sequence of constrained prior volumes. As higher-likelihood regions are explored, the evidence estimate converges, providing a principled termination criterion.

The \textsc{nessai} framework accelerates this process by learning the distribution of the region of parameter space with likelihood above the current threshold. It achieves this using normalizing flows—neural-network-based models capable of representing complex, non-Gaussian distributions—which greatly improve sampling efficiency and reduce the computational cost of nested sampling.

Once the sampler has converged, we obtain both the posterior distribution and a robust estimate of the evidence for the hypothesis. These evidence values become the numerator and denominator of the Bayes factor, thereby quantifying how well each waveform model explains the observed data. The second term on the RHS of Eq.~\eqref{eq:odds_ratio}, known as the \textit{prior ratio}, weights the BF by the relative prior probabilities of the two hypotheses.

In this work we consider the distinguishability between BBH waveforms lensed by CSs and (i) unlensed waveforms, and (ii) waveforms lensed by a PML. For the first case, assuming that the generation of a detectable BBH signal and its lensing are independent events, the prior ratio becomes 
\begin{equation}\label{eq:prior_ratio_independent}
    \frac{p(\mathcal{H}_{\rm CS})}{p(\mathcal{H}_U)}=\frac{p_{\rm CS}\;p_{\text{BBH}}}{p_{\text{BBH}}}=p_{\rm CS}\,,
\end{equation}
where $\mathcal{H}_{\rm U}$ denotes the unlensed hypothesis. The lensing probability can be estimated as
\begin{equation}\label{eq:lensing_p}
    p_{\rm CS} = \frac{N_{\rm LS}}{N_{\rm US} + N_{\rm LS}}\simeq\frac{N_{\rm LS}}{N_{\rm US}}\,,
\end{equation}
where $N_{\rm LS}$ is the number of detected lensed signals per year and $N_{\rm US}$ is the number of unlensed detected signals per year. The approximation assumes $N_{\rm LS}\ll N_{\rm US}$. Substituting $N_{\rm LS}$ with a central value of our calculated detection rate for CS-lensed signals (Appendix~\ref{sec-detection-rate}) ($\sim10^{-5}$) and $N_{\rm US}$ with the O4a detection rate (128 detections in 298 days, corresponding to $\sim156.7$~events/year) \cite{gwtc4}, we obtain an estimate for the prior probability as $\log(p_{\rm CS}) \approx -7.2^{-5.2}_{-9.2}$ depending on our assumption on the value of the $\Delta$ parameter.

For the comparison between CS-lensed and PML-lensed waveforms, the prior ratio is similarly given by 
\begin{equation}
    \label{eq:prior_ratio_pml_cs}
    \frac{p(\mathcal{H}_{\rm CS})}{p(\mathcal{H}_{\rm PML})} = \frac{p_{\rm CS}\;p_{\rm BBH}}{p_{\rm PML}\;p_{\rm BBH}} =  \frac{p_{\rm CS}}{p_{\rm PML}}\,.
\end{equation}
Using detection ratios, we estimate 
\begin{equation}
    \label{eq:p_csl_over_p_pml}
    \frac{p_{\rm CS}}{p_{\rm PML}} = \frac{(N_S + N_{\rm PML})N_{\rm CS}}{(N_S + N_{\rm CS})N_{\rm PML}}\simeq\frac{N_{\rm CS}}{N_{\rm PML}}\,,
\end{equation}
assuming again that $N_{\rm CS}, N_{\rm PML} \ll N_S$. 

To grossly estimate this prior ratio we will use the constraints on the strong lensing rate  from \cite{lensingO3b}, considering only the detection of single images, since these could already carry a microlensing imprint. By using these values we are assuming that strong lensing induced by galaxies or galaxy clusters also induce microlensing signatures. We note that this assumption produces an overestimation of the microlensing detection rate, as this might not be the case. For our analysis we will use, as a first-order approximation, that the microlensing induced by galaxies and galaxy clusters can be modeled by the PML transmission factor and we will use a central value for the strong lensing rate of $\sim10^{-3}$. We can estimate two bounds of the prior rate by looking at the two extreme cases: (i) the case with highest value of $\Delta$ and lower lensing rate and (ii) the case with lowest $\Delta$ and highest lensing rate. This produces prior ratios in the following regime: $\log(\frac{p_{\rm CS}}{p_{\rm PML}})\approx -4.2^{-1.5}_{-6.5}$. 

Bayesian model selection is highly sensitive to the SNR of the detected signal. Higher SNRs allow parameter estimation algorithms to extract more detailed features, improving model discrimination. 
Conversely, low-SNR signals may obscure key features that would give away information about the particular source of the signal. To quantify this effect, we compute the odds ratio for three BBH waveform injections, using the prelensing SNRs listed in Table~\ref{tab:waveform_parameters}. The used priors for the estimation of the BF are provided in Table~\ref{tab:bf_priors}.

\begin{table}[b!]
\caption{Priors used for the estimation of the Bayes factor. For the BBH parameters we used the default priors provided by {\sc Bilby}, while the priors for the PML and CS lensing parameters were defined within the range where the beating pattern is visible in the frequency range of interest.}
\label{tab:bf_priors}
\vspace{2mm}
\centering
\begin{tabular}{c|c}
\hline
Parameter & Prior \\
\hline
\hline
\multicolumn{2}{c}{BBH parameters}\\
\hline
$\mathcal{M}$ & Uniform $[15, 100] \;\rm M_\odot$ \\\hline
$q=M_2/M_1$ & Uniform\footnote{Individual masses are constrained between 5 and 100 M$_\odot$.} $[0.125, 1]$ \\\hline
$D_L$ & Uniform\footnote{This prior is uniform in the source frame.} $[100, 8000]\,$Mpc \\
\hline
\multicolumn{2}{c}{CS lensing parameters}\\\hline
%$t_\Delta$ & Uniform $[5\cdot10^{-4}, 1.3]\,$s \\
$t_\Delta$ & Uniform\footnote{The parameter $\Delta$ is not sampled independently, but is derived from $t_\Delta$ via Eq.~\eqref{t_Delta}, yielding $\Delta \in [7\times10^{-11},\,3.5\times10^{-9}]$ for $\chi_L=1\,$Gpc. We note that $\Delta \propto \chi_L^{-1/2}$ and thus decreases with increasing $\chi_L$.} $[5\cdot10^{-4}, 1.3]\,$s \\
$y$ & Uniform $[5\cdot10^{-4}, 1.3]$\\
\hline
\multicolumn{2}{c}{PML lensing parameters}\\\hline
$t_M$ & Uniform $[5\cdot10^{-4}, 1.3]\,$s \\
$y$ & Uniform $[10^{-2}, 1.3]$\\\hline
\end{tabular}
\end{table}

The results of this comparison are compiled in Fig.~\ref{fig:detectability}, where the distribution of the odds ratio across the lensing parameters for each injection are shown. Taking logarithms in Eq.~\eqref{eq:odds_ratio} a log-odds ratio of $x$ implies that one hypothesis is $x$ orders of magnitude more probable than the other, given the data. The plots show that CS lensing signatures are generally distinguishable from both unlensed and PML-lensed hypotheses, although not uniformly across the entire lensing parameter space. This variation across injections is expected.

Characteristic features of the CS transmission factor fall within the detector's sensitivity band roughly when $t_\Delta>0.01\,$s, as seen in the log-odds ratio plots, where amplification peaks and oscillations appear in the relevant frequency range. The behavior with respect to the impact parameter $y$ remains relatively consistent across the explored region.

%%%%%%%%%%%%%%%%%%%%%%%%%%%%%%%%%%%%%%%%%%%%%%%%%%%

%--------------------------------------------------
\section{Conclusions}
%--------------------------------------------------
\label{sec-concl}

In this study, we introduced a framework tailored for detecting gravitational lensing by cosmic strings, addressing an important gap in current gravitational-wave searches. The method is based on the full-wave transmission factor $F$, which captures the diffraction and interference patterns generated by the conical spacetime geometry of a string and can be directly employed for efficient template generation.  
This solution is computationally efficient, expressed as a sum of two standard Fresnel integrals readily available in scientific libraries such as Python’s {\sc scipy}, and significantly faster than analogous full-wave treatments for compact mass lenses—such as the point mass lens,  which involves hypergeometric functions, or the singular isothermal sphere  model, which requires infinite series expansions.
This makes it straightforward to incorporate the transmission factor into existing LVK data-analysis pipelines for searching cosmic string signatures. 

A central aspect of our work is the direct comparison between CS lensing and the standard PML model. The differences are fundamental and observationally significant. PML lensing is governed by a localized gravitational potential, admits the thin-lens approximation, and yields images determined by a lens equation. By contrast, CS lensing is topological, arising from the global conical structure of spacetime, which invalidates a strict thin-lens treatment. PMLs produce two images with magnifications and phase shifts set by the potential, whereas CSs generate two identical non-amplified images with no Morse phase shift.
The parameter controlling wave effects in a PML is the distance-independent ratio $2R_{\rm S}/\lambda$, while for a CS it is $\mathcal{D}\Delta^2/\lambda$, which grows with the lens-observer distance $\mathcal{D}$. These differences lead to distinct waveform morphologies: microlensing appears as a beating pattern from overlapping images, while strong lensing yields two time-shifted replicas with delay $\Delta t_{21} = 2y\,t_{\Delta}$. We derived a detectability bound on the string tension, 
Eq.~\eqref{eq:Delta_max},
which sets a benchmark for searches.  

We also quantified the selection bias in searching for CS-lensed signals with unlensed template banks. The frequency modulation induced by $F(f)$ reduces the matched-filter SNR, with the largest losses occurring when the beating frequency $f_{\Delta}$ falls within the LVK band ($20$–$1000$ Hz). The $\chi^2$ signal-consistency test introduces an additional downranking, contributing up to $\sim 60\%$ of the SNR loss in extreme cases. The mismatch peaks at intermediate frequencies where detector sensitivity is greatest, amplifying the impact of lensing. Finally, Bayesian model selection shows that CS-lensed signals are distinguishable from both unlensed and PML-lensed hypotheses across a wide range of parameters. This distinctiveness, rooted in the characteristic phase and amplitude structure encoded in $F$, demonstrates that targeted searches can successfully identify CS lensing.  
It would also be interesting to explore the overlap between CS lensing and the phenomenological search for single distorted signals~\cite{LiuA-Wong23}, which similarly produces beating patterns like those of PML and CS lensing. Assessing the distinguishability of CS-lensed signals in that broader context could provide further insight into their observational signatures. While this is beyond the scope of the present work, it represents a natural direction for future investigation.

An important aspect of our analysis concerns the role of the string tension $\Delta$ and the lens distance $\chi_L$ in shaping observational signatures and their interplay with source parameter degeneracies. The lensing time delay $t_\Delta \propto \chi_L \Delta^2$ [Eq.~\eqref{t_Delta}] directly controls the frequency scale of interference fringes, which can mimic intrinsic features of the binary waveform---such as spin-induced modulations or eccentricity---if not properly modeled. This highlights the need for dedicated template families that incorporate cosmic string lensing, as standard unlensed templates may misattribute lensing-induced modulations to astrophysical parameters.
%Breaking these degeneracies will likely require joint inference on lensing and source parameters, leveraging multi-band observations or complementary electromagnetic constraints.}

Finally, Eqs.~\eqref{eq:lower-bound}--\eqref{eq:Delta_max_LVK} quantify the detectability threshold for lensing signatures. Enforcing $f_{00} < f_+$ imposes a lower bound on the time delay $t_\Delta$, and hence on the string tension $\Delta$, which scales as
\begin{equation}
\Delta_{\min} \propto (\chi_L f_+)^{-1/2}.
\end{equation}
This dependence on the product of lens distance and detector frequency band introduces an important trade-off: while low-frequency detectors such as LISA require larger tensions for nearby lenses, this requirement can be partially compensated by cosmological distances, where $\chi_L$ is large and the time delay grows accordingly. For LVK ($f_+ \sim 10^3$ Hz), the threshold is $\Delta \gtrsim 10^{-10}$ for lenses at $\sim 100$ Mpc, whereas next-generation ground-based detectors with higher $f_+$ will probe even smaller tensions. Conversely, LISA operating at millihertz frequencies would need $\Delta \gtrsim 10^{-8}$ for nearby lenses, but could detect weaker strings if they lie at gigaparsec scales. This frequency-distance interplay underscores the complementarity between detector bands: terrestrial interferometers are optimal for lensing by weak strings at moderate distances, while space-based missions target strong strings or distant lenses, alongside their sensitivity to direct GW emission from string networks.

Overall, our framework  offers a starting point for dedicated searches, enhancing the prospects of discovering cosmic strings through gravitational-wave lensing observations and opening a new window onto high-energy physics. 

\begin{acknowledgements}
The authors are grateful to Mick Wright for his helpful comments on the manuscript. 
This work was supported by the Spanish Ministry of Science and Innovation through grants PID2021-125485NB-C21, PID2021-125485NB-C22, PID2024-159689NB-C21, and PID2024-159689NB-C22 funded by MCIN/AEI/10.13039/501100011033 and the European Regional Development Fund (ERDF), ‘A way of making Europe’. O.B.~acknowledges 
financial support from grant CEX2024-001451-M, funded by MICIU/AEI/10.13039/501100011033, and 
support from the Agència de Gestió d’Ajuts Universitaris i de Recerca (AGAUR), Generalitat de Catalunya (grant SGR-2021-01069). N.V.~and J.A.F.~are supported by the Generalitat Valenciana (grant CIPROM/2022/49). J.A.F.~acknowledges further support from the European Horizon Europe staff exchange (SE) programme HORIZON-MSCA2021-SE-01 Grant No.~NewFunFiCO-101086251. The authors gratefully acknowledge the computer resources provided by the Nyx cluster at the ICCUB and by Artemisa, as well as the technical support from the Instituto de Física Corpuscular (IFIC, CSIC-UV). Artemisa is cofunded by the European Union through the 2014-2020 ERDF Operational Programme of the Comunitat Valenciana under project IDIFEDER/2018/048.
This material is based upon work supported by NSF’s LIGO Laboratory which is a major facility fully funded by the National Science Foundation.
\end{acknowledgements}

%%%%%%%%%%%%%%%%%%%%%%%%%%%%%%%%%%%%%%%%%%%%%%%

\appendix

\section{Characterization of Nodal and Antinodal Structures}
\label{appendixA}

\begin{figure}[h!]
\includegraphics[width=0.9\columnwidth]{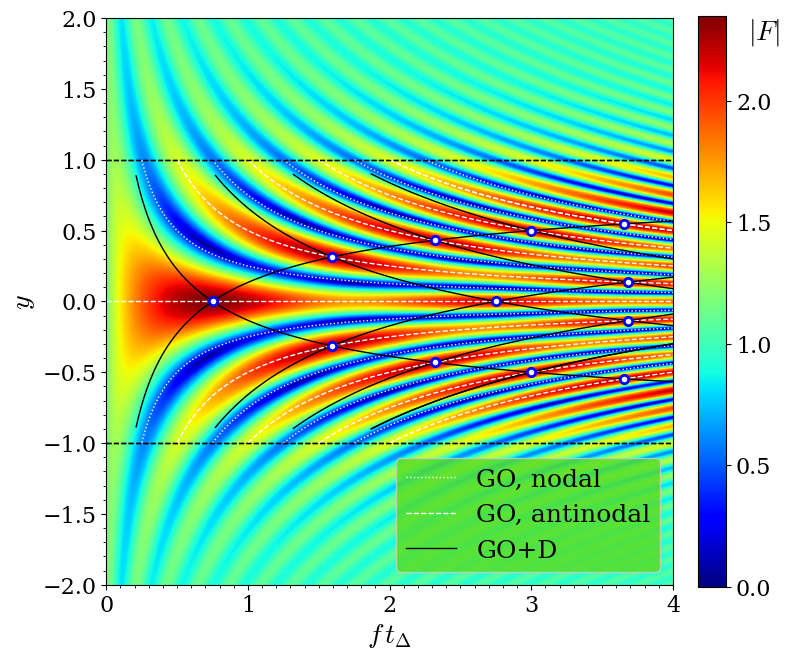}
\caption{Lines of constant phase overlaid on the wave pattern of the full-wave solution of Fig.~\ref{fig:F-string-full}. Circles mark the positions of the local diffraction maxima determined by Eqs~\eqref{eq:fnm} and \eqref{eq:ynm}.
The solid black lines show the curves connecting the maxima for a fixed $m$, as described by Eq.~\eqref{eq:first-diffract}.
}
\label{fig:noadal-lines}
\end{figure}

In Fig.~\ref{fig:F-string-full}, wave effects appear as interference fringes in the transmission factor across parameter space. Inside the double-imaging region ($|y| < 1$), these fringes are accurately captured by the interference between the two GO rays [Fig.~\ref{fig:F-approx}(a)]. To quantify this behavior, we note that the maxima and minima arise when the phase difference satisfies
\begin{equation}
2\pi f \,(t_d^+ - t_d^-) = \pi q
\end{equation}
where $q$ is an integer. This condition leads to hyperbolic curves in the $(f,y)$ space, as depicted in Fig.~\ref{fig:noadal-lines}, and can be expressed as
\begin{equation}
y = \displaystyle{\frac{q}{4ft_{\Delta}}}, \quad \text{with}\; q=
\begin{cases}
0, \pm 2, \dots \; \text{in antinodal lines,} \\ 
\pm 1, \pm 3, \dots \; \text{in nodal lines.}
\end{cases}
\label{eq:go-interf}
\end{equation}

Including the third, diffracted contribution in Eq.~\eqref{eq:F-GTD} accounts for further structure in the interference pattern: namely, the local amplification maxima along the antinodal lines and the appearance of fringes in the shadow region ($|y|>1$), where one of the GO rays is absent.
The diffraction term acquires a phase shift of $3\pi/4$, as evident in the diffraction coefficients \eqref{eq:Diff}. 
This leads to the condition for amplification maxima due to interference between GO and diffracted waves \cite{Jopt-mm18}
\begin{equation}
2f \,t_d^+ = n + \frac{3}{4}, \qquad
2f \,t_d^- = m + \frac{3}{4},
\end{equation}
where $n, m$ are even integers: $n,m= 0, 2, 4, \dots$.
This leads to amplification maxima located at
\begin{equation}
f_{nm}   = \frac{1}{4\,t_{\Delta}} \left(\sqrt{n +\frac{3}{4}} + \sqrt{m +\frac{3}{4}}\right)^2,
\label{eq:fnm}
\end{equation}
\begin{equation}
y_{nm} = \frac{n - m}{4 f_{nm}\,t_{\Delta}}.
\label{eq:ynm}
\end{equation}
As an illustration, the strongest diffraction maximum lies on the line of sight ($y=0$), obtained for $m$=$n$=$0$, yielding the frequency $f_{00}= 3/(4\,t_{\Delta})$.
The next-order maxima appears at frequency 
$f_{20}=(\sqrt{11}+\sqrt{3})^2/(16t_{\Delta}) \approx 1.6/t_{\Delta}$, 
and $y_{20}=(\sqrt{11}-\sqrt{3})/(\sqrt{11}+\sqrt{3})\approx 0.314$.
These values agree well with the pattern shown in Fig.~\ref{fig:noadal-lines}.
Due to symmetry, the relations $f_{mn}=f_{nm}$ and $y_{mn}=-y_{nm}$ hold.
The diffraction maxima can be connected by a continuous line (shown as solid black lines in the figure) for a fixed index $m$ and varying index $n$. The curves are determined by the following formula: 
\beq
y_m(f)=\pm \left[ 1-\sqrt{(m+3/4)/(f t_{\Delta}) } \right]
\label{eq:first-diffract}
\eeq
with $m= 0, 2, 4, \dots$.

%%%%%%%%%%%%%%%  Appendix B  %%%%%%%%%%%%%%%%

\section{Line-of-sight approximation}
\label{line-of-sight}

Of particular interest is the case when the string lies close to the line of sight ($y \ll 1$).
To leading order, the relative time delay between the two GO rays can be neglected, so that $t_d^+ = t_d^- = t_{\Delta}/2$.
From Eq.\ \eqref{eq:F-f-full}, the transmission factor becomes
\begin{equation}
F_0(f) =2\eee{-\ii \pi ft_{\Delta}} \mathcal{F}(\sqrt{\pi ft_{\Delta}}) \approx 2\eee{-\ii \pi ft_{\Delta}} -  \frac{\eee{\ii \pi/4}}{\pi\sqrt{ft_{\Delta}}}. 
\end{equation}
The corresponding absolute value is
\begin{equation}
|F_0(f)| \approx 2 \left[1 
- \frac{1}{\pi\sqrt{ft_{\Delta}}} \cos\left(\pi ft_{\Delta} + \frac{\pi}{4}\right)
\right]^{1/2},
\label{eq:F0}
\end{equation}
where terms smaller than $O(f^{-1/2})$ are neglected.
Equation~\eqref{eq:F0} describes decaying oscillations around the asymptotic value $|F_0|=2$, corresponding to the GO limit.  
These oscillations are due to diffraction effects.  
The strongest maximum occurs at frequency $f_{00}\,t_{\Delta} = 3/4$, yielding the amplification
\beq
|F|_{max}= 2\, [1+2/(\pi\sqrt{3})]^{1/2} \approx 2.34
\label{eq:F_max}
\eeq
in agreement with Fig.~4 of Ref.~\cite{pla-string16} (noting differences in notation).

It is worth noting that amplification by a factor of 2 is a distinctive feature of gravitational lensing by a CS. Due to the conical topology of the spacetime, the source is effectively duplicated, producing two identical images that interfere coherently. 
At first glance, this resembles Young’s double-slit experiment, where interference fringes scale as $\sim \lambda r/d$, with $d$ the slit separation~\cite{born-wolf-03}. However, the analogy is only superficial. Unlike the double-slit configuration, a conical spacetime lacks an intrinsic length scale. Instead, the interference and diffraction are governed solely by the deficit angle $\Delta$, resulting in behaviors qualitatively distinct from those of the classical optical setup~\cite{Jopt-mm18}.

%%%%%%%%%%%%%%%%%%%%%%%%%%%%%%%%%%%%%%%%%%%%

\section{Detection rate}
\label{sec-detection-rate}

In this appendix, we describe how we compute the lensing detection rate for cosmic strings. We consider a network of three O3 detectors and impose three detection criteria. First, the unlensed waveform must be detectable with a sufficiently large signal-to-noise ratio (in this work, we require $SNR\geq 8$). Second, the lensing-induced interference fringes must be statistically significant, which we enforce by requiring a mismatch of at least $0.05$ between the lensed and unlensed waveforms. Third, the parameter $\Delta$ must satisfy the condition given in Eq.~\eqref{eq:Delta_max}. Taken together, these criteria ensure that both the signal and its lensing imprint are detectable and distinguishable. Now, to compute the detection rate of lensed waveforms, we integrate over all possible source binaries and cosmic string configurations that satisfy the above criteria. The resulting expression for the detection rate is given by \cite{jung18} 
\begin{equation}
    \text{detection rate}\,(\Delta, M) \;=\; n_S(M)\, n_L\, V(\Delta, M),
    \label{eq:c1}
\end{equation}
where $n_L$ is the density of cosmic strings per comoving volume, $n_S(M)$ is the merger rate density per year and per comoving volume as a function of the total mass of the binary system, $M=M_1+M_2$, and $V(\Delta, M)$ is a six-dimensional volume in the space of string and source locations that yields detectable lensing. This $V$ accounts for all spatial configurations of one string and one source that produce an interference fringe satisfying the detection criteria. Accordingly, $V(\Delta, M)$ has units of (comoving volume)$^2$ and effectively counts all source-string configurations (for a given $M$ and $\Delta$) that would be observable as a  lensed fringe event. This volume factorizes as 
\begin{eqnarray}\label{eq:Vdelta}
    V(\Delta,M) \;=\; \int_{\chi_{\min}}^{\chi_{\max}} 4\pi\,\chi_L^2 \;V_S(\chi_L,\Delta,M)\;d\chi_L,
    \label{eq:c2}
\end{eqnarray}
where $V_S(\chi_L,\Delta,M)$ is the 3D volume of source locations, for binaries of total mass $M$, that produce detectable fringes when a string of tension $\Delta$ is located at comoving distance $\chi_L$. The integration over $\chi_L$ runs from the minimum comoving distance to the string $\chi_{\min}(\Delta)$ determined by Eq.~\eqref{eq:Delta_max}, up to the detector's horizon, $\chi_{\max}(M)$. 

For a given string at $\chi_L$, several conditions limit $V_S$. First, the source must lie behind the string (so that the string is between the source and the observer). If $\chi_S$ denotes the source comoving distance, then this requires $\chi_S > \chi_L$. Second, the source must be within the detector horizon: $\chi_S \le \chi_{\max}(M)$, where $\chi_{\max}(M)$ is the maximum comoving distance at which a binary of mass $M$ can be detected with $SNR \geq 8$. This $\chi_{\max}$ is essentially set by the SNR threshold and corresponds to the unlensed detection range for that mass. Finally, the relative alignment of the source with respect to the string must produce observable interference fringes. This alignment is quantified by the dimensionless impact parameter $y=\theta/\Delta$, where $\theta$ is the angular separation between the source and the line of sight passing through the string, evaluated at the string position (see also Sec.~\ref{subsec:full-wave}, where the parameter $y$ is introduced in the expression for the transmission factor).
In this parametrization, $y = 0$ corresponds to a source located along the line of sight directly behind the string, while $y = 1$ corresponds to a source lying at the boundary of the double-imaging region, set by the deficit angle $\Delta$, which approximately marks the extent of the string’s gravitational influence. We define $y_{\max}$ as the maximum impact parameter for which the fringe detectability conditions (mismatch and resolvability) are satisfied. This $y_{\max}$ then determines the maximum angular separation between the source and the line of sight as measured at the observer, $\phi_{\max}$:
\begin{equation}
    \phi_{\max}=\text{arcsin}\bigg[\frac{d_{LS}}{d_S}\sin (y_{\max}\,\Delta)\bigg],
\end{equation}
where $d_{LS}=(\chi_S-\chi_L)/(1+z_S)$ is the angular diameter distance between the lens and the source \cite{urrutia21}. Since the angle $\theta_{\max}=y_{\max}\,\Delta$ is very small (of the order of $\Delta$), we can approximate $\phi_{\max}$ as 
\begin{equation}\label{eq:phimax}
    \phi_{\max} \approx \frac{\chi_S-\chi_L}{\chi_S} \,y_{\max}\,\Delta \approx \bigg(1-\frac{\chi_L}{\chi_S}\bigg)\,y_{\max}\,\Delta,
\end{equation}
where we have used $d_S=\chi_S/(1+z_S)$. For given $\chi_L$ and $\chi_S$, the area of the sphere of radius $\chi_S$ containing the sources that satisfy the criteria can then be approximated by $\phi_{\max}\,\chi_S \, L(\chi_L,\chi_S)$, where $L$ is the length of the geodesic connecting the two points at which the cosmic string intersects the sphere of radius $\chi_S$:
\begin{equation}\label{eq:L_int}
L(\chi_L,\chi_S)=2\chi_S \, 
\text{arcsin}\sqrt{1-\bigg(\frac{\chi_L}{\chi_S}\bigg)^2}
\end{equation}
By integrating over all source distances from $\chi_L$ up to the detector horizon $\chi_{\max}(M)$, we obtain the source volume corresponding to a string located at $\chi_L$: 
\begin{equation}
    V_S(\chi_L,\Delta,M) = \int_{\chi_L}^{\chi_{\max}} 
       \phi_{\max} \,L(\chi_L,\chi_S)\,\chi_S \, d\chi_S \, ,
\end{equation}
where $\phi_{\max}(\chi_S,\chi_L,M,\Delta)$ is given by Eq.~\eqref{eq:phimax} and $L(\chi_L,\chi_S)$ by Eq.~\eqref{eq:L_int}.

\begin{figure}[h!]
    \centering
    \includegraphics[width=1\linewidth]{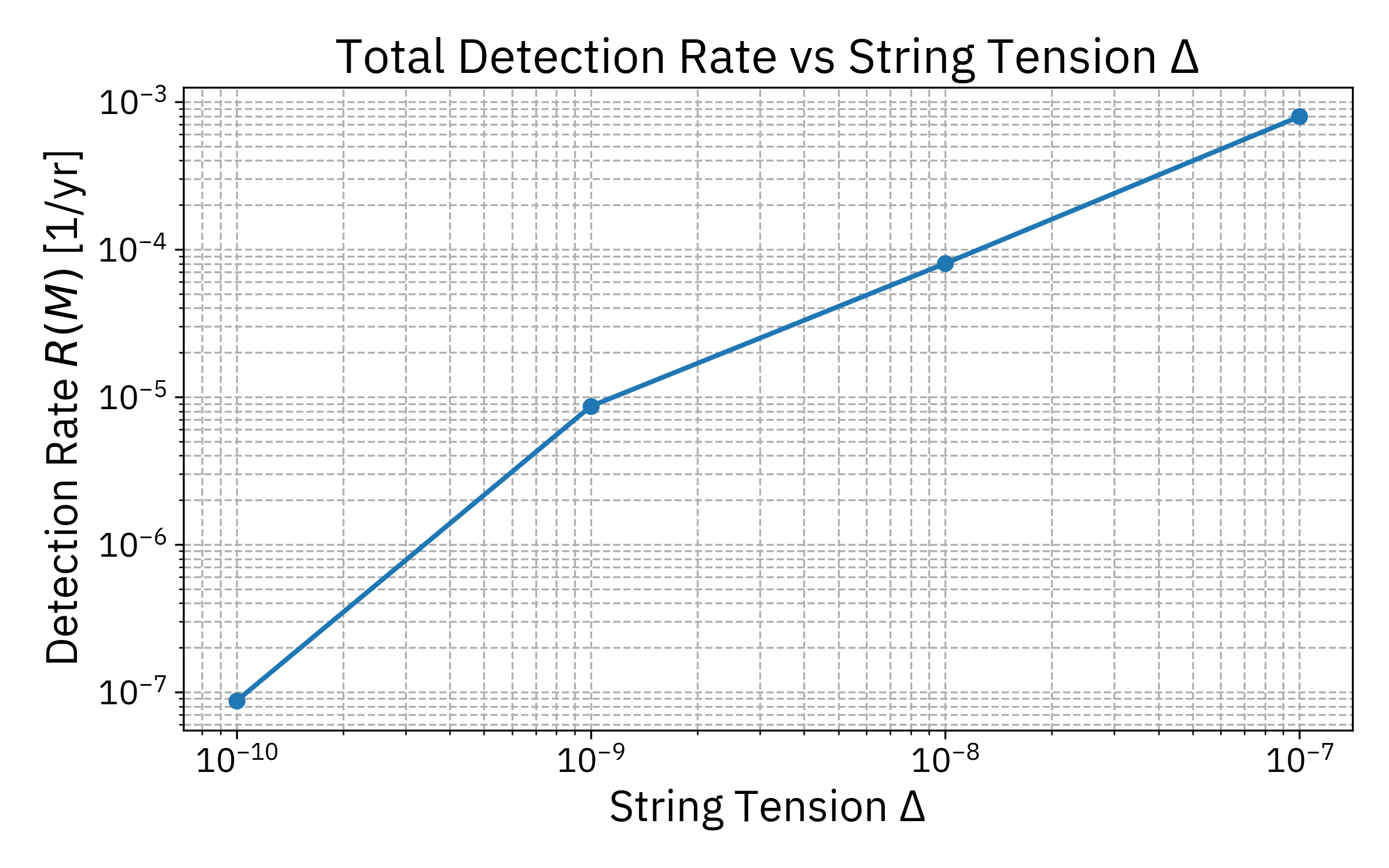}
    \caption{Total detection rate for different of $\Delta$.}
    \label{fig:total-detection-rate}
\end{figure}

The overall detection rate is finally obtained by integrating Eq.~\eqref{eq:c1} over the binary mass distribution: 
%\begin{equation} 
\begin{align} \label{eq:Rdelta}
R(\Delta) 
&= \int dM\, n_S(M)\, n_L\, V(\Delta,M) \nonumber \\
&= 8\pi n_L \Delta \, \int dM \, n_S(M) \int d\chi_S \int d\chi_L \, y_{\max} 
\nonumber \\
&\times \chi_L^2 \,\chi_S^2 \, \left(1 - \frac{\chi_L}{\chi_S}\right) \,
\arcsin\!\sqrt{1 - \left(\frac{\chi_L}{\chi_S}\right)^2} .
\end{align}
%\end{equation}
The dependence on the string tension $\Delta$ enters not only through the overall prefactor, but also implicitly via the integration limits on $\chi_L$ [see Eq.~\eqref{eq:Vdelta}]. In practice, the integrals are evaluated by summing over discrete bins in mass and distance. We adopt an optimistic astrophysical merger-rate model for $n_S(M)$ (the M10 model) and assume a constant cosmic-string density $n_L = N_L / V_H$, where $N_L$ denotes the number of strings within a Hubble volume $V_H$. 
Following Ref.~\cite{Xiao2022FRBStrings}, $N_L$ is estimated using the relation $n_L=N_L(z)H(z)^3$, which effectively counts the number of strings per Hubble volume. Using the parametrization developed in \cite{Buschmann_2022}, Xiao et al.~\cite{Xiao2022FRBStrings} find that, for an underlying phase transition occurring at $T \simeq 10^{15},\mathrm{GeV}$, the present-day network contains approximately $N_L(z=0) \simeq 30$ strings per Hubble volume. We adopt this value as our fiducial choice. Nevertheless, we note that other studies favor a smaller, order-unity number of long strings per Hubble volume, $N_L \sim \mathcal{O}(1)$ \cite{Gorghetto_2018,Hindmarsh_2020}. Our constraints  scale linearly with $N_L$: adopting a different number of strings per Hubble volume would rescale the predicted lensing rates by $N_L/30$ but would not qualitatively change our conclusions. The results of these calculations are shown in Fig.~\ref{fig:total-detection-rate}.
We observe that the detection rates decrease for small values of $\Delta$. This behavior arises because the condition imposed by Eq.~\eqref{eq:Delta_max} excludes the closest lenses, corresponding to small values of $\chi_L$, from contributing to the rate.
%%%%%%%%%%%%%%%%%%%%%%%%%%%%%%%%%%%%%%%%%%

%\bibliographystyle{apsrev4-2}
%\newpage
\bibliography{references}

\end{document}